\documentstyle[nato,epsf]{crckapb} 

\begin{opening}
\title{THE STANDARD MODEL}
\author{M. HERRERO}
\institute{Departamento de Fisica Teorica\\
           Facultad de Ciencias, C-XI\\
           Universidad Autonoma de Madrid\\
           Cantoblanco, 28049 Madrid, Spain\\
           e-mail:herrero@delta.ft.uam.es}
\end{opening}
 
\runningtitle{THE STANDARD MODEL }

\begin{document}
\begin{center}
 ABSTRACT
\end{center}
\begin{quotation}
\noindent
 These lectures provide an introduction to the basic aspects of the 
 Standard Model, $SU(3)_{C} \times SU(2)_{L} \times U(1)_{Y}$.
 \footnote{Lectures presented at the NATO ASI 98 School, Techniques and
 Concepts of High Energy Physics; St. Croix, Virgin Islands, USA, June 18-29
 1998\\
 {}\\
{\large FTUAM 98/25}\\
{Dec.-1998}\\
{hep-ph/9812242} }
\end{quotation}
 

\def\co{SU(3)_{C}}
\def\ws{SU(2)_{L} \times U(1)_{Y}}
\def\group{SU(3)_{C} \times SU(2)_{L} \times U(1)_{Y}} 
\def\em{U(1)_{em}}
\def\lsim{\mathrel{\lower2.5pt\vbox{\lineskip=0pt\baselineskip=0pt
\hbox{$<$}\hbox{$\sim$}}}}
\def\gsim{\mathrel{\lower2.5pt\vbox{\lineskip=0pt\baselineskip=0pt
\hbox{$>$}\hbox{$\sim$}}}}
\def\gs{SU(2)_{\rm L} \times U(1)_{\rm Y}}
\def\lr{SU(2)_L \times SU(2)_R}
\def\lpr{SU(2)_{L+R}}
\def\sbs{{\cal SBS}}
\def\wh{{\cal W}}
\def\bh{{\cal B}}
\def\wtu{\wh^{\mu \nu}}
\def\wtd{\wh_{\mu \nu}}
\def\btu{\bh^{\mu \nu}}
\def\btd{\bh_{\mu \nu}}
\def\cl{{\cal L}}
\def\nll{\cl_{\rm NL}}
\def\ecl{\cl_{\rm EChL}}
\def\fpnl{\cl_{\rm FP}^{\rm NL}}
\def\fpl{\cl_{\rm FP}}
\def\msb{{\overline{\rm MS}}}
\def\mh{M_H} 

\section{Introduction}


All known particle physics phenomena  are extremely well described within
the Santard Model (SM) of elementary particles and their fundamental 
interactions. The SM provides a very elegant theoretical framework and it has 
succesfully passed very precise tests which at present are at the $0.1\%$
level \cite{gws,books,qcd,status}.

We understand by elementary particles the point-like constituents of matter with no
known substructure  up to the present limits of $10^{-18}-10^{-19}m$. These are of two
types, the basic building blocks of matter themselves konwn as matter particles
 and the intermediate interaction particles. The first ones are fermions of spin 
 $s=\frac{1}{2}$ and are classified into leptons and quarks. The known leptons are: the
 electron, $e^-$, the muon, $\mu^-$ and the $\tau^-$ with electric charge $Q=-1$ (all
 charges are given in units
 of the elementary charge $e$); and the corresponding neutrinos $\nu_e$, $\nu_{\mu}$ and
 $\nu_{\tau}$ with $Q=0$. The known quarks are of six different flavors: $u$, $d$, $s$,
 $c$, $b$ and $t$ and have fractional charge $Q=$ $\frac{2}{3}$,$-\frac{1}{3}$,
 $-\frac{1}{3}$,
 $\frac{2}{3}$, $-\frac{1}{3}$ and $\frac{2}{3}$ respectively.
 
  The
 quarks have an additional quantum number, the color, which for them can be of three
 types, generically denoted as $q_i$, $i=1,2,3$. We know that color is not seen in Nature 
 and
 therefore the elementary quarks must be confined into the experimentally observed
 matter particles, the hadrons.  These colorless composite particles are classified
 into baryons and mesons. The baryons  are fermions made of three quarks, $qqq$, 
 as for instance the proton, $p\sim uud$, and the neutron, $n\sim ddu$.  
  The mesons are bosons made of one
  quark and one antiquark as for instance the pions, $\pi^+\sim u{\bar d}$ and 
 $\pi^-\sim d{\bar u}$.
 
 The second kind of elementary particles are the intermediate interaction
 particles. By leaving apart the gravitational interactions, 
 all the relevant interactions in Particle Physics are known to be 
 mediated by the
 exchange of an elementary particle that is a boson with spin $s=1$. 
 The photon, $\gamma$, is the exchanged particle in the electromagnetic
 interactions, the eight gluons $g_{\alpha}\;;\alpha=1,..8$ mediate the strong
 interactions among quarks, and the three weak bosons, $W^{\pm}$, $Z$ are the
 corresponding intermediate bosons of the weak interactions.
 
 As for the theoretical aspects, the SM is a quantum field theory 
 that is based on the  gauge symmetry $\group$. 
  This  
 gauge group includes the symmetry group of the strong interactions, $ \co$,
 and the symmetry group of the electroweak interactions,
 $\ws$.   
   The group symmetry of the electromagnetic interactions, $ \em$,
 appears in the SM as a subgroup of $ \ws$ and it
 is in
 this sense that the weak and electromagnetic interactions are said to be
 unified. 
 
 {\bf The gauge sector} of the SM 
   is composed  of eight gluons which are the gauge bosons of $ \co$ and the 
 $\gamma$, $W^{\pm}$ and $Z$ particles which are the four gauge bosons of 
 $ \ws$.
 The main physical properties of these intermediate gauge bosons 
  are as follows. The gluons are massless, electrically neutral and carry color
  quantum number. There are eight gluons since they come in eight different
  colors. The consequence of the gluons being colorful is that they interact not
  just with the quarks but also with themselves. The weak bosons, $W^{\pm}$ and
  $Z$ are massive particles and also selfinteracting. The $W^{\pm}$ are
  charged with $Q=\pm 1$ respectively and the $Z$ is electricaly neutral. The
  photon $\gamma$ is massless, chargeless and non-selfinteracting.  
 
 Concerning the range of the various interactions, it is well known 
 the infinite range of the electromagnetic interactions as it corresponds to
 an interaction mediated by a massless gauge boson, the short range of the weak
 interactions of about $10^{-16}cm$ correspondingly to the exchange of a
 massive gauge particle with a mass of the order of $M_V\sim 100\, GeV$ and, finally, the
 strong interactions whose range is not infinite, as it should correspond to the
 exchange of a massless gluon, but finite due to the extra physical
 property of confinement. In fact, the short range of the strong interactions
 of about $10^{-13}cm$ corresponds to the typical size of the ligthest
 hadrons.  
 
 As for the strength of the three interactions, the electromagnetic
 interactions are  governed by the size of the electromagnetic coupling
 constant $e$ or equivalently $\alpha=\frac{e^2}{4\pi}$ which at low energies
 is given by the fine structure constant, $\alpha(Q=m_e) =\frac {1}{137}$.
 The weak interactions at energies much lower than the exchanged gauge boson
 mass, $M_V$, have an effective (weak) strength given by the dimensionful 
 Fermi constant $G_F=1.167\times 10^{-5}\, GeV^{-2}$. The name of strong
 interactions is due to their comparative stronger strength than the other
 interactions. This strength is governed by the size of the
 strong copling constant $g_S$ or equivalently $\alpha_S=\frac{g_s^2}{4\pi}$
 and is varies from large values to low energies,
  $\alpha_S(Q=m_{\rm hadron}) \sim 1$ up to the vanishing asymptotic limit 
  $\alpha_S(Q\rightarrow\infty)\rightarrow 0$. This last limit 
  indicates that
  the quarks behave as free particles when they are observed at infinitely
  large energies or, equivalently, inflinitely short distances and it is known
  as the property of asymptotic freedom. 
  
 Finally, regarding the present status of the matter particle content of 
 the SM the situation is summarized as follows.
 
  {\bf The fermionic
 sector } of quarks and leptons are organized in three families  with
 identical properties except for mass. The particle content in each family is:
 \vspace{1cm}
 \begin{center}
 $1^{st}$ family:  
$\left(\begin{array}{c}
 \nu_e \\
  e^- 
\end{array} \right)_L$, $e^-_R$,  
 $\left(\begin{array}{c}
 u \\
 d  
\end{array} \right)_L$, $u_R$, $d_R$ 
\end{center} 
 
\begin{center}
 $2^{nd}$ family:  
$\left(\begin{array}{c}
 \nu_{\mu} \\
  \mu^- 
\end{array} \right)_L$, $\mu^-_R$,  
 $\left(\begin{array}{c}
 c \\
 s  
\end{array} \right)_L$, $c_R$, $s_R$ 
\end{center} 
 
\begin{center}
 $3^{rd}$ family:  
$\left(\begin{array}{c}
 \nu_{\tau} \\
  \tau^- 
\end{array} \right)_L$, $\tau^-_R$,  
 $\left(\begin{array}{c}
 t \\
 b  
\end{array} \right)_L$, $t_R$, $b_R$ 
\end{center} 
\vspace{1cm}
and their corresponding antiparticles. The left-handed and right-handed fields
are defined by means of the chirality operator $\gamma_5$ as usual,
\vspace{0.5cm}
\begin{center}
$e_L^-=\frac{1}{2}(1-\gamma_5)e^-$; $e_R^-=\frac{1}{2}(1+\gamma_5)e^-$      
\end{center}
\vspace{0.5cm}
and they transform as doublets and singlets of $SU(2)_L$ respectively. 

{\bf The scalar sector} of the SM is not experimentaly confirmed
yet. The fact that the weak gauge bosons are massive particles, $M_W^{\pm}$,
$M_Z \ne 0$, indicates that $\ws$ is {\it NOT} a symmetry of the vacuum. In
contrast, the photon being massless reflects that $\em$ is a good symmetry of
the vacuum.  Therefore, the Spontaneous Symmetry Breaking pattern in the
SM must be:
\vspace{0.5cm}
\begin{center}
$\group \rightarrow \co\times\em$
\end{center}
\vspace{0.5cm}
The above pattern is implemented in the SM by means of the
so-called Higgs Mechanism which provides the proper masses to the $W^\pm$ and
$Z$ gauge bosons and to the fermions, and 
leaves as a consequence the prediction of a
new particle: The Higgs boson particle. This must be scalar and electrically
neutral. This particle has not been seen in the experiments so far \cite{sbs}. 

These lectures provide an introduction to the basic aspects of the SM,
$\group$. They aim to be of pedagogical character and are especially addressed 
to non-expert particle physicists without much theoretical background in
Quantum Field Theory. The lectures start with a review on some symmetry
concepts that are relevant in particle physics, with particular emphasis in the
concept of gauge symmetry. Quantum Electrodynamics (QED) is introduced next 
as a paradigmatic example of
gauge theory. A short review on the most relevant precursors of Quantum
Chromodynamics (QCD) and the
Electroweak Theory are presented. Then, a brief introduction to the basics
of QCD is presented. The central part of these lectures is devoted to the
building of the Electroweak Theory and to review the Electroweak Symmetry
Breaking in the SM.  The concept of Spontaneous Symmetry Breaking and The Higgs
Mechanism are explained. A short review on the present theoretical Higgs mass
bounds is also included. The final part of these lectures is devoted to review
the most relevant SM predictions.

There are many other important aspects of the SM that, due to the lack of space
and time, are not covered here. Some of the complementary topics
are covered by my fellow lecturers at this School. In particular, the lectures 
of J. Stirling cover 
QCD, those of R. Aleksan cover Quark Mixing and CP Violation and those of L.
Nodulman cover the Experimental Tests of the SM.

\section{Group Symmetries in Particle Physics }

The existence of symmetries plays a crutial role in Particle Physics.
We say that there is a symmetry $S$ when the physical stystem under study has an
invariance under the transformation given by $S$ or, equivalently, when the
Hamiltonian of this system $H$ is invariant, i.e., 
\begin{center}
$SHS^+=H$
\end{center}
Sometimes the set of independent symmetries of a system generates an algebraic
structure of a group, in which case it is said there is a symmetry 
group \cite{ChengLi}.

\subsection{Types of symmetries}
There are various ways of classifying the different symmetries. If we pay
attention to the kind of parameters defining these symmetry transformations,
they can be classified into:
\begin{itemize}
\item[1.-] {\it Discrete Symmetries}\\
The parameters can take just discrete values. In Particle Physics there are
several examples. Among the most relevant ones are the transformations of:
\begin{center}
Parity $P$, Charge Conjugation $C$ and Time Reversal $T$
\end{center}
On the other hand, by the $CPT$ Theorem we know that all interactions must 
be invariant under the total transformation given by the three of them $C$, $P$
and $T$, irrespectively of their order. It is also known that the
elctromagnetic interactions and the strong interactions preserve in addition 
$P$, $C$ and
$T$ separately, whereas the weak interactions can violate, $P$, $C$ and $PC$.
\item[2.-] {\it Continous Symmetries}\\
The parameters take continous values. The typical examples are the rotations
,generically written as $R(\theta)$, where the rotation angle $\theta$ can take
continous values. There are different kinds of continous symmetries. Here we
mention two types: 
\subitem 1)  
{\it Space-Time symmetries}: Symmetries that act on the space-time. Typical
examples are traslations, rotations, etc. 
\subitem 2) {\it Internal Symmetries}: Symmetries that act on the internal quantum
numbers. Typical examples are $SU(2)$ Isospin symmetry, $U(1)_B$ 
baryon symmetry etc. Usually these symmetries are given by Lie groups. 
\end{itemize}
\subsection{Irreducible representations of a symmetry group}  
 The classification of the particle spectra into irreducible 
 representations of a given symmetry group is an important aspect of Particle
 Physics and helps in understanding their basic physical properties. 
 
 In the case of rotations, it is known that if a system given by a
 particle of spin $j$ manifests $SO(3)$ 
 rotational invariance, then it implies the existence of $(2j+1)$ degenerate
 energy levels which can be accommodated into the irreducible representation            
  of $SO(3)$ of dimension $(2j+1)$. Let us see this in more detail.
  
 Let $R(\theta)$ be a the rotation given by:
 $$\displaystyle  R(\theta)=e^{i\displaystyle 
  \sum^3_{a=1}\theta_a J_a}\; ;\;J_a={\rm angular\; momentum\;
  operators} $$
   Invariance of the Hamiltonian under $R(\theta)$ implies the following sequence
 of statements, 
 \vspace{0.3cm} 
 \begin{center}
 $RHR^+=H \Longrightarrow [H,J_a]=0 \;(a=1,2,3) \Longrightarrow 
 {\rm If}\; \{|n>\}/H|n>=E_n|n> \; {\rm Then}\; 
 H(J_a|n>)=J_a(H|n>)=E_n(J_a|n>)$
 \end{center}
 \vspace{0.3cm}
 Thus, there are $(2j+1)$ states, the $J_a|n>$, that are degenerate
 and form the basis associated to angular momentum $j$. In addition,
 the angular momentum operators $J_a$ are the generators of the symmetry group 
 of rotations $SO(3)$ which is the set of all the $3\times 3$ orthogonal
 matrices with unit determinant. By putting altogether, the conclusion is that 
  the $SO(3)$ symmetry of $H$
 implies that each particle with angular momentum $j$ has $(2j+1)$ degenerate
 levels which fit into the irreducible representation with dimension $(2j+1)$ 
 of the $SO(3)$ group.   
 
\subsection{Internal Symmetries}
The Internal Symmetries are transformations not on the space-time 
coordinates but on internal coordinates, and they transform  one 
particle to another with different internal quantum numbers but having the same
mass. In contrast to the case of space-time symmetries, the irreducible
representations of internal symmetries are degenerate particle multiplets.
\subsubsection{$SU(2)$ Isospin Symmetry}
The isospin symmetry is an illustrative example of internal symmetries. In this
case the internal quantum number is isospin. Let us see that, in fact,
invariance under isospin implies the existence of degenerate isospin
multiplets.  
Let $H_s$ be the Hamiltonian of strong interactions. 
Invariance of strong interactions under isospin rotations reads:
\begin{center}
$UH_sU^+=H_s$
\end{center}
where, $U$ is the isospin transformation and is given by
$$ \displaystyle U=e^{i\displaystyle \sum^3_{a=1}\theta_aT_a}$$
with $T_a$ ($a=1,2,3$) being the three generators of the $SU(2)$ group and 
$\theta_a$ the continous parameters of the transformation. The $SU(2)$ group 
is the set
of $2\times 2$ unitary matrices with unit determinant; and the $SU(2)$ algebra
is defined by the conmutation relations of the generators:
\begin{center}
$\left[ T_i,T_j \right] \;=\;\epsilon_{ijk}T_k$; $\epsilon_{ijk}={\rm
structure\;constants\;of}\;SU(2)$
\end{center}
As in the previous case of space-time transformations, one can show that
invariance under the above $U$ transformation implies,
\begin{center}
$\left[ T_a,H_s \right] =0$; $(a=1,2,3)$
\end{center}
and from this it is immediate to demonstrate the existence of degenerate
isospin multiplets. Thus, for a given eigenstate of $H_s$ one can always find, by
application of the $T_a$ generators, new eigenstates of $H_s$ which are
degenerate.  

In terms of the physical states, the proton $|p>$, the neutron $|n>$, and the
pions, $|\pi^+>$, $|\pi^->$ and $|\pi^0>$ the isospin rotations act as follows:

\begin{center}
$T_+|n>=|p>$; $T_-|p>=|n>$\\
$T_3|p>=\frac{1}{2}|p>$; $T_3|n>=-\frac{1}{2}|n>$
\end{center}
\begin{center}
$T_+|\pi^->=\sqrt{2}|\pi^0>$; $T_+|\pi^0>=\sqrt{2}|\pi^+>$\\
$T_3|\pi^+>=|\pi^+>$; $T_3|\pi^0>=0$; $T_3|\pi^->=-|\pi^->$
\end{center}

where, $T_{\pm}=T_1 \pm iT_2$.

Therefore the corresponding degenerate isospin multiplets are:
\begin{center}
\framebox{$\left (
\begin{array}{c}
p\\n 
\end{array}\right )$ isospin doublet}\hspace{0.2cm} 
\framebox{$\left (
\begin{array}{c}
\pi^+\\ \pi^0\\ \pi^- 
\end{array}\right )$ isospin triplet}
\end{center}

Notice that neither the proton and neutron nor the three pions are exactly
degenerate and therefore the isospin symmetry is not an exact symmetry of the
strong interactions. In fact, the size of the mass-differences within
 a multiplet are 
indications of the size of the isospin breaking. Since the proton and 
neutron masses are pretty close and simmilarly for the masses of the three pions, it
happens that the $SU(2)$ isospin symmetry is, indeed, an approximate 
symmetry of the
strong interactions.
   
\subsection{ Classification of internal symmetries and relevant theorems}
There are two distinct classes of internal symmetries:
\begin{itemize}
\item[1.-] {\it Global symmetries}\\
The continous parameters of the transformation {\it DO NOT DEPEND} on the
space-time coordinates. Some examples are: $SU(2)$
 Isospin symmetry, $SU(3)$ flavor symmetry, $U(1)_B$ baryon symmetry, 
 $U(1)_L$ lepton symmetry,...
 \item[2.-] {\it Local (Gauge) symmetries}\\
 The continous parameters of the transformation {\it DO DEPEND} on the
 space-time coordinates. Some examples are: $U(1)_{em}$ electromagnetic
 symmetry, $SU(2)_L$ weak isospin symmetry, $U(1)_Y$ weak hypercharge symmetry,
 $SU(3)_C$ color symmetry,...
 \end{itemize}
 There are two relevant theorems/principles that apply to the two cases above 
 respectively and have important physical implications:\\

\vspace{0.3 cm}
\underline {Noether's Theorem for Global Symmetries}\\
\vspace{0.1cm}
 If the Hamiltonian (or the Lagrangian) of a physical system 
 has a global symmetry, there must be  
a current and the associated charge that are conserved.\\
\begin{itemize}
\item[] 
{\it Examples:}\\
The $U(1)$ symmetries are global rotations by a given phase. For instance:
\begin{center}
$\Psi \rightarrow e^{i\alpha} \Psi$
\end{center}
rotates the field $\Psi$ by a phase $\alpha$ and it is the same for all space
time points, i.e. it is a global phase. The $U(1)$ symmetry group is the set of
complex numbers with unit modulus. We have already mentioned some examples in
particle physics as the $U(1)_B$ and $U(1)_L$ global symmetries. The conserved
currents are the barionic and leptonic currents respectively; and the
associated conserved charges are the barionic and leptonic numbers
respectively. 
\end{itemize}
\vspace{0.3cm}
\underline {The Gauge Principle for Gauge Theories}\\ 
\vspace{0.1cm}
Let $\Psi$ be a physical system in Particle Physics whose dynamics is decribed 
by a Lagrangian ${\cal L}$ which is invariant under a global symmetry $G$. 
It turns out that, by  promoting this global symmetry $G$ from global to local,
the originaly free theory transforms into an interacting theory. The
procedure in order to get the theory invariant under local transformations
is by introducing new vector boson fields, the so-called gauge fields, that
interact with the $\Psi$ field in a gauge invariant manner. The number of gauge
fields and the particular form of these gauge invariant interactions depend on
the particularities of the symmetry group $G$. More specifically, the number of associated
gauge boson fields is equal to the number of generators of the symmetry group
$G$.
\begin{itemize}
\item[] 
{\it Examples:}\\ 
 The local version
of the previous example, 

\begin{center}
$\Psi \rightarrow e^{i\alpha (x)} \Psi$
\end{center}
with the phase $\alpha$ being a function of the space-time point
$x\equiv x_{\mu}$, 
has one associated gauge boson field. This simplest case of $U(1)$ 
has just one generator and
correspondingly one gauge field which is the exchanged boson particle and acts
as the mediator of the corresponding interaction.\\
Other examples are: $SU(2)$ with three generators and the corresponding three
gauge bosons and $SU(3)$ with eight generators and the corresponding eight
gauge bosons. The generic case of $SU(N)$ has $N^2-1$ generators and
correspondingly the same number of gauge bosons.   
\end{itemize}

\vspace{0.3 cm} 
The above Gauge Principle is a very important aspect of 
 Particle Physics and has played a crutial role in the building of the
Standard Model.

The quantum field theories that are based on the existence 
of some gauge symmetry are called Gauge Theories. 
 We have already mentioned the cases of $U(1)_{em}$, $SU(2)_L$, $U(1)_Y$ and 
 $SU(3)_C$ gauge symmetries. The gauge theory based on $U(1)_{em}$ is  
 Quantum Electrodynamics (QED), the gauge theory based on $SU(3)_C$ is Quantum
 Chromodynamics (QCD) and the corresponding one based on the composed group
  $SU(2)_L \times U(1)_Y$ is the so-called Electroweak Theory. The Standard
  Model is the gauge theory based on the total gauge symmetry of the
  fundamental interactions in particle physics, $SU(3)_C \times SU(2)_L \times
  U(1)_Y$.

\section{QED: The paradigm of Gauge Theories} 
Quantum Electrodynamics is the most succesful Gauge Theory in Particle Physics
and  has been tested up to an extremely high level of precision. We show QED
here as the paradigmatic example of the application of The Gauge Principle and 
its physical implications. 

One starts with the following physical system:  A free Dirac field 
$\Psi$ with spin
$s=\frac{1}{2}$, mass $m$ and electric charge $Qe$. 

The corresponding Lagrangian is:

\begin{center}
${\cal L}= \overline{\Psi}(x)(i \partial{\hspace{-6pt}\slash}-m)\Psi(x)$; 
$\partial{\hspace{-6pt}\slash}\equiv \partial_{\mu}\gamma^{\mu}$
\end{center}

and the corresponding equation of motion is the Dirac equation:

\begin{center}
 $(i \partial{\hspace{-6pt}\slash}-m)\Psi(x)=0$
\end{center}
     
It is immediate to show the invariance of this  Lagrangian
 under global $U(1)$ transformations which act on
the fields and their derivatives as follows,

\begin{center}
$\Psi \rightarrow e^{iQ\theta}\Psi$; 
$\overline{\Psi}\rightarrow \overline{\Psi} e^{-iQ\theta}$;  
$\partial_{\mu}\Psi \rightarrow e^{iQ\theta}\partial_{\mu}\Psi$
\end{center}

here the global phase is $Q\theta$ and the continous papameter is $\theta$.

By Noether's Theorem, this global $U(1)$ invariance of ${\cal L}$ implies
the conservation of the electromagnetic current, $J_{\mu}$, 
and the electromagnetic charge, $eQ$,

\begin{center}
$J_{\mu}=\overline{\Psi}\gamma_{\mu}eQ\Psi$; $\partial_{\mu}J^{\mu}=0$;
 $eQ=\int d^3xJ_0(x)$
\end{center}

Now, if we promote the transformation from global to local, i.e, if the
parameter $\theta$ is allowed to depend on the space-time point $x$, 
the corresponding transformations on the fields and their derivatives are,

\begin{center}
$\Psi \rightarrow e^{iQ\theta (x)}\Psi$; 
$\overline{\Psi}\rightarrow \overline{\Psi} e^{-iQ\theta(x)}$;  
$\partial_{\mu}\Psi \rightarrow e^{iQ\theta(x)}\partial_{\mu}\Psi$ +
    $iQ(\partial_{\mu}\theta(x))e^{iQ\theta (x)}\Psi$
\end{center}   

One can show that the Lagrangian in the form written above is not yet 
invariant under these local transformations. The solution to this question is
provided by the Gauge Principle.  One introduces one gauge vector boson field,
the photon field $A_{\mu}(x)$ which interacts with the field $\Psi$ and 
transforms properly under the $U(1)$ gauge transformations,

\begin{center}
$A_{\mu}\rightarrow A_{\mu}-\frac{1}{e}\partial_{\mu}\theta (x)$
\end{center}

Here proper transformations means that it must compensate the extra terms
introduced by $\partial_{\mu}\theta\neq 0$ such that the total Lagrangian be 
finally gauge invariant.

The most economical way of building this gauge invariant Lagrangian is by
simply replacing the normal derivative, $\partial_{\mu}$, by the 
so-called covariant derivative, $D_{\mu}$, 

\begin{center}
$D_{\mu}\Psi\equiv (\partial_{\mu}-ieQA_{\mu})\Psi$
\end{center}

which transforms covariantly, i.e. as the $\Psi$ field itself,  

\begin{center}
$D_{\mu}\Psi \rightarrow e^{iQ\theta (x)}D_{\mu}\Psi$
\end{center} 

Finally, in order to include the propagation of the photon field one adds the
so-called kinetic term which must be also gauge invariant and is given in terms
of the field strength tensor,

\begin{center}
$F_{\mu\nu}=\partial_{\mu}A_{\nu}-\partial_{\nu}A_{\mu}$
\end{center} 

The total Lagrangian is Lorentz and U(1) gauge invariant, and is the well known
Lagrangian of QED,

\begin{center}
\framebox{${\cal L}_{QED}=\overline{\Psi}(x)
(i D{\hspace{-6pt}\slash}-m)\Psi(x)-\frac{1}{4}F_{\mu\nu}(x)F^{\mu\nu}(x)$}
\end{center}

Notice that it contains the wanted interactions within the
$\overline{\Psi}iD{\hspace{-6pt}\slash}\Psi$ term,

\begin{center}
$ \overline{\Psi}eQ A_{\mu}\gamma^{\mu}\Psi$
\end{center}
 
Finally, the gauge group for electromagnetism is, correspondingly, 
$U(1)_{em}$ with one generator, $Q$ and one parameter $\theta$.

\section{Strong Interactions before QCD: The Quark Model}

The discovery of the $\Lambda^0$ and $K^0$ particles lead to the proposal of a
new additive quantum number named 'strangeness' and denoted by $S$ which is
conserved by the strong interactions but is violated by the weak interactions.

The corresponding assignements are:

\begin{center}
$S(\Lambda^0)=-1$, $S(K^0)=+1$, $S(p)=S(n)=S(\pi)=0$.
\end{center}

Gell-Mann and independently Nishijima and Nakano by studing the properties of
the hadrons noted in 1953  the  
linear relation among the three additive quantum numbers the strangeness $S$,
the electromagnetic charge $Q$ and the third component of the (strong) isospin
$T_3$, given by the so-called 
{\it Gell-Mann-Nishijima relation} \cite{GellNish}:

\begin{center}
\framebox{$Q=T_3+\frac{Y}{2}$}
\end{center}

where the (strong) hypercharge $Y$, 
the baryon number $B$ and the 
strangeness $S$
are related by $Y=B+S$. 
 
The existence of the new conserved quantum number $S$ suggested  to think 
of a larger
symmetry than isospin $SU(2)$ for the strong interactions. Gell- Mann and 
Ne'eman in 1961 proposed  the larger symmetry group $SU(3)$, named sometimes 
flavor symmetry, which in fact contains to $SU(2)$ \cite{GellNee}. 
They pointed out that all
mesons and baryons with the same spin and parity can be grouped into
irreducible representations of $SU(3)$. Thus, each particle is labelled by its 
($T_3$, $Y$) quantum numbers and fits into one of the elements of 
these representations. Historically, it was named 'The Eightfold-Way'
classification of hadrons since the first studied hadrons turned out to fit
into representations of dimension eight, i.e. into octects of $SU(3)$. Later,
higher dimensional representations, as decuplets etc., were 
needed to fit other hadrons. In Figure 1
some examples of $SU(3)$ octects are shown.  

In 1964 Gell-Mann and Zweig \cite{GellZwe} noted that the lowest dimensional irreducible 
representation, i.e. the triplet of $SU(3)$ with dimension equal to three, 
 was not occupied by any known hadron and proposed the existence of
new particles, named {\it quarks}, such that by fitting them into the
elements of this fundamental representation  and by making
apropriate compositions  with it,
one could build up the whole
spectra of hadrons. This brilliant idea, originaly based mainly on formal
aspects of symmetries, led to the prediction of three new 
elementary  particles, the three lightest quarks, distinguished by their
flavors, the $u$ (up), $d$ (down) and $s$ (strange) quarks. Correspondingly,
their antiparticles, the antiquarks $\bar{u}$, $\bar{d}$ and $\bar{s}$ with the
opposite quantum numbers, are
fitted into the complex conjugate representation which is also a triplet. 
In Figure 1, these $SU(3)$ triplets are also shown.
 
\begin{figure}
\hspace{1.5cm}
\epsfysize8cm
\epsffile{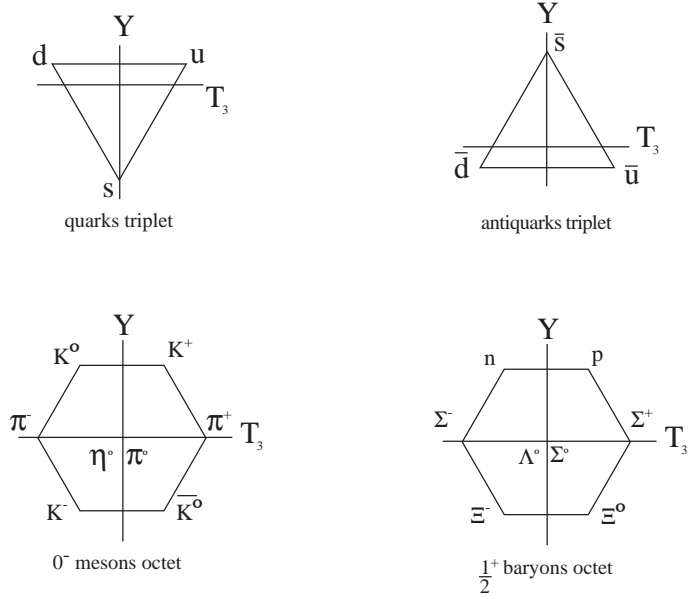}
\caption{Examples of particle multiplets in the $SU(3)$ Quark Model}
\end{figure}

The description of hadrons in terms of quarks by means of the $SU(3)$
irreducible representations and their properties is called the $SU(3)$ 
{\it Quark Model}. One uses group theory methods, for instance the Young
Tableaux technique, to decompose products of irreducible representations into
sums. Thus the mesons ($B=0$) appear as composite states of a quark and an 
antiquark, and the baryons ($B=1$) as composite states of three quarks:

\begin{center}
\underline{Mesons}: $q\bar{q}=3\otimes \bar{3}=1\oplus 8$
\end{center}

Here the 1 are the $SU(3)$ meson singlets as the $\eta '$ $\sim$  
$(u\bar{u}+d\bar{d}+s\bar{s})$ with $J^P=0^-$, the
$\Phi$ $\sim$ $(u\bar{u}+d\bar{d}+s\bar{s})$ 
with $J^P=1^-$... The 8 are the $SU(3)$ meson octets as the $0^-$
mesons, $\pi^+$ $\sim$ $u\bar{d}$, $K^+$ $\sim$ $u\bar{s}$
...;  the $1^-$ mesons, $\rho^+$ $\sim$ $u\bar{d}$, 
$K^{*+}$ $\sim$ $u\bar{s}$..., and so on.   

\begin{center}
\underline{Baryons}: $qqq=3\otimes 3 \otimes 3 = 
1\oplus 8 \oplus 8 \oplus 8 \oplus 10$
\end{center}
 
Here the 1 are the  $SU(3)$ baryon singlets; the 8 are the $SU(3)$ baryon octets 
as the $\frac{1}{2}^+$ baryons, $n$ $\sim$ $udd$, $p$ $\sim$ $uud$... ; the 10
are the $SU(3)$ decuplets as the $\frac{3}{2}^+$ baryons, $N^{*+}$ $\sim$
$uud$,  $\Sigma^{*+}$ $\sim$ $suu$..., and so on.

Notice that the $SU(3)$ flavor symmetry is not an exact symmetry. As in the
case of isospin symmetry, the mass differences within the members of a
multiplet are signals of $SU(3)$ breaking. Similarly, the mass differences among
the three quarks themselves are indications of this breaking as well. Although
the breaking is certainly more sizeable in $SU(3)$ than in $SU(2)$, one can
still deal with $SU(3)$ as an approximate symmetry. 

Among the successful predictions of the Quark Model there are, for instance, 
the existence of some particles before their discovery as it is the case of the
$\frac{3}{2}^+$ baryon $\Omega^-$ $\sim$ $sss$; and, in general, a large amount
of the hadron properties are well described within this model. In particular, 
one may
build up the hadron wave functions and compute some physical properties as, for
instance, the hadron magnetic moments, in terms of the corresponding ones of the
quarks components.   

For example, the proton wave function with the spin up is:

\begin{center}
$|p\uparrow>=\sqrt{\frac{1}{18}}|u\uparrow u\downarrow d\uparrow +  
 u\downarrow u\uparrow d\uparrow-2u\uparrow u\uparrow d\downarrow +perm.>$
\end{center}

The predicted proton and neutron magnetic moments are:  
     
\begin{center}
$\mu_p=\frac{4}{3}\mu_u-\frac{1}{3}\mu_d$;
 $\mu_n=\frac{4}{3}\mu_d-\frac{1}{3}\mu_u$; with, $\mu_q=\frac{Q_qe}{2m_q}$
\end{center}

and their ratio is, therefore,

\begin{center}
\framebox{$\frac{\mu_n}{\mu_p}=-\frac{2}{3}$}
\end{center}

which is rather close to its experimental measurement,

\begin{center}     
\framebox{$\left. \frac{\mu_n}{\mu_p}\right|_{exp}=-0.68497945\pm 0.00000058$}
\end{center}

\vspace{0.1cm}
The Quark Model has some noticeable failures which were the reason to 
abandon it as the proper model for strong interactions. The most famous one, 
from
the historial point of view, is the so-called {\it paradox of the} $\Delta^{++}$.
Its wave function for the case of spin up is given by,

\begin{center}
$|\Delta^{++} \uparrow >\sim |u \uparrow u \uparrow u \uparrow >$
\end{center}

which is apparently totally symmetric since it is symmetric in space, flavor
and spin. However, by Fermi-Dirac statistics it should be antisymmetric as it
corresponds to an state with identical fermions. This apparent paradox, was
solved by Gell-Mann with the proposal of the quarks carrying a new quantum
number, {\it the color} and, in consequence, being  non-identical 
from each other. Correspondingly, a quark $q$ can have 
three different colors, generically, $q_i\; i=1,2,3$.
Since, this property of color is not seen in Nature, the
colors of the quarks must be combined such that they produce colorless hadrons.
In the group theory language it is got by requiring the hadrons to be in the
singlet representations of the color group, $SU(3)_C$. Since the singlet
representation is allways antisymmetric, by including this color wave function 
one gets finally the expected
antisymmetry of the $\Delta^{++}$ total wave function.

\section{QCD: The Gauge Theory of Strong Interactions}


Quantum Chromodynamics is the gauge theory for strong interactions and has
provided plenty of successful predictions so far \cite{qcd}. 

It is based on the gauge
symmetry of strong interactions, i.e. the local color transformations which 
leave its Hamiltonian (or Lagrangian) invariant. 
The gauge symmetry group that is generated by these color transformations is
the non-abelian Lie group $SU(3)_C$. Here $C$ refers to colors and 3 refers to
the three posible color states of the quarks which are assumed to be in the
fundamental representation of the group having dimension three. The gluons are
the gauge boson particles associated to this gauge symmetry and are eight of
them as it corresponds to the number of $SU(3)$ generators. The gluons are the
mediators of the strong interactions among quarks. Genericaly, the quarks 
and gluons are denoted
by:

\begin{center}
\underline{quarks}: $q_i$, $i=1,2,3$; \hspace{1cm}\underline{gluons}: 
$g_{\alpha}$, $\alpha=1,...,8$
\end{center}

The building of the QCD invariant Lagrangian is done by following the same
steps as in the QED case. In particular, one applies the gauge principle as
well with the particularities of the non-abelian group $SU(3)$ taken into
account. Thus, the global symmetry $SU(3)$ of the Lagrangian for the strong
interactions is promoted to local by  replacing the derivative of the quark
by  its covariant derivative which in the QCD case is,

\begin{center}
$D_{\mu}q\equiv \left( \partial_{\mu}-ig_s
(\frac{\lambda_{\alpha}}{2})A_{\mu}^{\alpha} \right) q$
\end{center}
  
where,

\begin{center}
$q=\left( \begin{array}{c} q_1\\q_2\\q_3 \end{array} \right )$
\end{center}

$$
\begin{array}{lcl}
q_i&=&\; {\rm quark}\; {\rm fields};\; 
i=1,2,3\\
g_s&=& \;{\rm strong}\; {\rm coupling}\; {\rm constant}\\
\frac{\lambda_{\alpha}}{2}&=&SU(3)\; {\rm generators} \\
A_{\mu}^{\alpha}&=&\; {\rm gluon}\;{\rm fields};\; \alpha=1,...,8
\end{array}
$$

The QCD Lagrangian is then written in terms of the quarks and their covariant
derivatives  and contains in addition the kinetic term for the gluon fields,
\begin{center}
\vspace{0.3cm}
\framebox{ $\displaystyle {\cal L}_{QCD}= \sum_{q} \overline{q}(x)
 (i D{\hspace{-6pt}\slash}-m_q)q(x) 
- \frac{1}{4}F_{\mu\nu}^{\alpha}(x)F^{\mu\nu}_{\alpha}(x)$  }
 \vspace{0.3cm}
\end{center}
The gluon field strength is,

\begin{center}
$F_{\mu\nu}^{\alpha}(x)=\partial_{\mu}A_{\nu}^{\alpha}(x)
  -\partial_{\nu}A_{\mu}^{\alpha}(x)
  +g_sf^{\alpha\beta\gamma}A_{\mu\beta}A_{\nu\gamma}$
\end{center}

and contains a bilinear term in the gluon fields as it corresponds to a
non-abelian gauge theory with structure constants $f^{\alpha\beta\gamma}$ 
($\alpha,\beta,\gamma=1,...,8$).

It can be shown that the above Lagrangian is invariant under the following 
$SU(3)$ gauge transformations,

\begin{center}
$\left\{
\begin{array}{lcl}    
 q(x)&\rightarrow & e^{i\theta_{\alpha}(x)\frac{\lambda^{\alpha}}{2}}q(x)\\
 D_{\mu}q(x)&\rightarrow &
 e^{i\theta_{\alpha}(x)\frac{\lambda^{\alpha}}{2}}D_{\mu}q(x)\\
 A_{\mu}^{\alpha}(x)&\rightarrow & A_{\mu}^{\alpha}(x)
 -\frac{1}{g_s}\partial_{\mu}\theta^{\alpha}(x)
 +f^{\alpha\beta\gamma}\theta_{\beta}(x)A_{\mu\gamma}(x) 
 \end{array}
 \right. $
\end{center}

where $\theta_{\alpha}(x)\; \alpha=1,...,8$ are the parameters of the 
transformation.
 
Similarly to the QED case, the gauge interactions among the quarks 
and gluons are contained in the $\bar{q}i D{\hspace{-6pt}\slash}q$ term,

\begin{center}
$\bar{q} g_s\frac{\lambda^{\alpha}}{2}A_{\mu}^{\alpha}\gamma^{\mu}q$
\end{center}

There is, however, an important difference  with the QED case. The gluon 
kinetic term $F_{\mu\nu}^{\alpha}F^{\mu\nu}_{\alpha}$ contains a three gluons
term and a four gluons term. These are precisely the selfinteraction gluon
vertices which are genuine of a non-abelian theory.  

\subsection{$SU(3)$ Group properties}

In this section we list the basic properties of the $SU(3)$ group which are 
relevant for QCD and in particular for the computation of color factors in 
processes mediated by strong interactions.

$SU(3)$ is the set of $3\times 3$ unitary matrices with unit determinant.
Any element of $SU(3)$, $U$, can be written in terms of its 8 generators,
$\frac{\lambda_{\alpha}}{2}$ and a set of 8 real parameters $\theta_{\alpha}$ 
as,

\begin{center}
$U=e^{i\theta^{\alpha}\frac{\lambda_{\alpha}}{2}}$; $\alpha=1,...,8$
\end{center}

The generators are $3\times 3$ traceless hermitian matrices 
$\frac{\lambda_{\alpha}}{2}$ and are given in 
terms of the so-called Gell-Mann matrices, $\lambda_{\alpha}$,

$$
\lambda_1=\left( \begin{array}{ccc}
0&1&0\\1&0&0\\0&0&0 \end{array} \right)\;\;,\;\;
\lambda_2=\left( \begin{array}{ccc}
0&-i&0\\i&0&0\\0&0&0 \end{array} \right)\;\;,\;\;
\lambda_3=\left( \begin{array}{ccc}
1&0&0\\0&-1&0\\0&0&0 \end{array} \right)    
$$

$$ 
\lambda_4=\left( \begin{array}{ccc}
0&0&1\\0&0&0\\1&0&0 \end{array} \right)\;\;,\;\;
\lambda_5=\left( \begin{array}{ccc}
0&0&-i\\0&0&0\\i&0&0 \end{array} \right) 
$$

$$   
\lambda_6=\left( \begin{array}{ccc}
0&0&0\\0&0&1\\0&1&0 \end{array} \right)\;\;,\;\;
\lambda_7=\left( \begin{array}{ccc}
0&0&0\\0&0&-i\\0&i&0 \end{array} \right)\;\;,\;\;
\lambda_8=\frac{1}{\sqrt{3}}\left( \begin{array}{ccc}
1&0&0\\0&1&0\\0&0&-2 \end{array} \right)    
$$

 Some basic properties of the $SU(3)$ generators are:

\begin{center}
\framebox{
$\displaystyle
\left[ \frac{\lambda_{\alpha}}{2},\frac{\lambda_{\beta}}{2}\right]
  \;=\; if_{\alpha\beta\gamma}\frac{\lambda_{\gamma}}{2}$ }\hspace{1cm}
\framebox{ 
$\displaystyle
Tr\left( \frac{\lambda_{\alpha}}{2}\frac{\lambda_{\beta}}{2}\right)
 \;=\;\frac{1}{2}\delta_{\alpha\beta}$}
\end{center}

The tensor $f_{\alpha\beta\gamma}$ is totally antisymmetric 
and its elements are the structure constants of $SU(3)$. 
The non-vanishing elements are,

\begin{center}
$f_{123}=1$, $f_{147}=\frac{1}{2}$, $f_{156}=-\frac{1}{2}$, 
$f_{246}=\frac{1}{2}$,
$f_{257}=\frac{1}{2}$,\\
$f_{345}=\frac{1}{2}$, $f_{367}=-\frac{1}{2}$, $f_{458}=\sqrt{\frac{3}{2}}$
$f_{678}=\sqrt{\frac{3}{2}}$
\end{center}

 Some useful relations 
for practical computations and relevant group factors are, 
 
$$\delta_{\alpha\beta}C_A\;=\;\sum_{\gamma\delta}f_{\alpha\gamma\delta}
 f_{\beta\gamma\delta}
\;\;;\;\;C_A=3$$

$$\delta_{ik}C_F\;=\;\sum_{\alpha l} 
\frac{\lambda^{\alpha}_{il}}{2}\frac{\lambda^{\alpha}_{lk}}{2} 
\;\;;\;\;C_F=\frac{4}{3}$$

$$\delta_{\alpha\beta}T_F\;=\;\sum_{ki}
\frac{\lambda^{\alpha}_{ik}}{2}\frac{\lambda^{\beta}_{ki}}{2}
\;\;;\;\;T_F=\frac{1}{2}$$

$$(i,k=1,2,3\;;\;\alpha,\beta,\gamma,\delta=1,...,8)$$

\subsection{Computing color factors in QCD}
For practical computations, sometimes it is convenient to define color
factors associated to a physical proccess in QCD. These color factors are
genuine of QCD and can be computed apart by using the $SU(3)$ group relations.

For illustrative purposes,
we present here one particular example: the computation 
of the color factor associated to the scattering proccess of two different
quarks, $qq'\rightarrow qq'$. 

There is just one Feynman diagram contributing to the scattering amplitude 
of this proccess  which is the diagram with one gluon exchanged in the
t-channel. Notice that this diagram is similar to the 
one contributing to the QED proccess $e^-\mu^-\rightarrow e^-\mu^-$
where one photon is exchanged in the t-channel. 

One can use the known result from QED for the spin-averaged squared 
amplitude of the proccess $e^-\mu^-\rightarrow e^-\mu^-$ in terms of
the Mandelstam variables, $s$, $t$, and $u$,

$$ \overline{|F|}^2=2e^4\left(\frac{s^2+u^2}{t^2}\right)$$

and by simply replacing the electromagnetic coupling constant $e$ by the strong
coupling constant $g_s$ and  by adding a color factor $F_C$ one obtains the 
corresponding squared
amplitude of the proccess $qq'\rightarrow qq'$, at tree level in QCD,

$$\overline{|F|}^2=F_C2g_s^4\left(\frac{s^2+u^2}{t^2}\right)$$
 
The color factor $F_C$ can be now computed apart. From the QCD Feynman rules
for the quark-gluon-quark vertex and for the gluon propagator; 
and by averaging in initial colors and summing
in final colors one gets,

$$F_C=\frac{1}{9}\sum_{ijlm\\ \alpha\beta\alpha'\beta'}
\left(\frac{\lambda_{ij}^{\alpha}}{2}\right)  
\delta_{\alpha\beta}  
\left(\frac{\lambda_{lm}^{\beta}}{2}\right)  
\left(\frac{\lambda_{ij}^{\alpha'}}{2}\right)^*
\delta_{\alpha'\beta'}
\left(\frac{\lambda_{lm}^{\beta'}}{2}\right)^*$$ 

The above expression can be simplified by using the  
properties of the $SU(3)$ generators,

$$\displaystyle
\begin{array}{lcl}
   F_C & = & \frac{1}{9} \sum_{\alpha\alpha'}
     \left[\sum_{ij}\left(\frac{\lambda_{ij}^{\alpha}}{2}\right) 
                    \left(\frac{\lambda_{ji}^{\alpha'}}{2}\right) \right]
      \left[\sum_{lm}\left(\frac{\lambda_{lm}^{\alpha}}{2}\right) 
                    \left(\frac{\lambda_{ml}^{\alpha'}}{2}\right) \right] \\
     &   &  \\                
     & = & \frac{1}{9} \sum_{\alpha\alpha'}
\left[\delta_{\alpha\alpha'}T_F\right]\left[\delta_{\alpha\alpha'}T_F\right]\\
     &   &   \\  
     & = & \frac{2}{9} 
\end{array}
$$

\section{Weak Interactions before The Electroweak Theory}

 The existence of new interactions of weak strength were proposed to explain 
 the experimental data indicating long lifetimes in the decays
 of known particles, as for instance,
 
$$n\rightarrow pe^-\overline{\nu}_e \;;\;  \tau_n=920\;sec$$ 
$$\pi^-\rightarrow \mu^- \overline{\nu}_{\mu}\;;\;\tau_{\pi^-}=2.6\times 10^{-8}\;sec$$ 
$$\mu^-\rightarrow e^-\overline{\nu}_e\nu_{\mu}\;;\;\tau_{\mu}=2.2\times 10^{-6}\;sec$$    
    
 These are much longer lifetimes than the typical decays mediated by strong  
 interactions  as,
 $$ \Delta \rightarrow p \pi \;;\; \tau_{\Delta}=10^{-23}\;sec$$
 and by electromagnetic interactions as,
 
 $$\pi^0 \rightarrow \gamma\gamma\;;\; \tau_{\pi^0}=10^{-16}\;sec$$  
      
The history of weak interactions before the formulation of the
  Standard $\ws$ Theory is an  
interesting example of the relevant interplay between theory and experiment.
There were a sequence of proposed models which were confronted systematically 
with the abundant 
experimental data and  which needed to be either refined or rejected  in
order to be compatible with the observations. All this relevant phenomenology 
of weak interactions together with the advent of the gauge theories
in particle physics led finally to
the formulation of the Electroweak Theory, i.e. the gauge theory of electroweak
interactions.   
Among the most relevant predecessor
theories of electroweak interactions are the following: 
Fermi Theory, V-A Theory of Feynman
and Gell-Mann and the IVB theory of Lee, Yang and Glashow.

\subsection{Fermi Theory of Weak Interactions}
In 1934 Fermi proposed the four-fermion interactions theory \cite{Fermi}
in order to
describe the neutron $\beta$-decay $n\rightarrow p e^- \overline{\nu}_e$,

\vspace{0.3cm}
\begin{center}
\framebox{
${\cal L}_F=-\frac{G_F}{\sqrt{2}}
\left [ \overline{p}(x)\gamma_{\lambda}n(x) \right ] 
\left [ \overline{e}(x)\gamma^{\lambda} \nu_e(x) \right ] + h.c. $   }
\end{center}
\vspace{0.3cm}
where the fermion field operators are denoted by their particle names and,

$$ G_F=1.167\times 10^{-5}\; GeV^{-2} $$
is the so-called Fermi constant which provides the effective dimensionful 
coupling of the weak interactions.

The Fermi Lagrangian above assumes a vector structure, as in the electromagnetic
 case, 
for both the hadronic
current, $J_{\lambda}^{(h)}(x)= \overline{p}(x)\gamma_{\lambda}n(x)$, 
and the leptonic current, 
$J_{\lambda}^{(l)}(x)= \overline{\nu}_e(x)\gamma_{\lambda}e(x)$; and postulates
a local character for the four fermion interactions, namely, the two currents
are contracted at the same space-time point $x$. 

Due precisely to the above vector structure of the weak currents, the Fermi
Lagrangian does not explain the observed parity violation in weak interactions.
 
\subsection{Parity Violation and the V-A form of charged weak interactions}
 The observation of Kaon decays in two different final states with opposite
 parities,
 $$ K^+\rightarrow \pi^+ \pi^0 \;\; {\rm and} \;\; 
 K^+\rightarrow \pi^+\pi^+\pi^- $$
 led Lee and Yang in 1956 to suggest the non-conservation of parity in  
 the weak interactions responsible for these decays \cite{LeeYang}. 
 Parity violation was
 discovered by Wu and collaborators in 1957 \cite{Wu}
 by  analizing the decays of 
 $Co$ nuclei 
 $$ ^{60}Co \rightarrow  ^{60}Ni^*\;e^-\;\overline{\nu}_e $$
 which proceed via neutron decay $n\rightarrow pe^-\overline{\nu}_e$. 
  
 The nuclei are polarized by the action of an external magnetic field such that
 the angular momenta for $Co$ and $Ni$ are $J=5$ and $J=4$ respectively, both
 aligned in the direction of the external field. By
 conservation of the total angular momentum, the angular momentum of
 the combined system 
 electron-antineutrino is inferred to be $J(e^-\;\overline{\nu}_e)=1$ and   
 must be  aligned  with the other angular momenta.
 Therefore both the electron and the
 antineutrino must have their spins polarized in this same direction. 
  The electron from the decay is seeing 
 always moving
 in the opposite direction to the external field.
 By total momentum conservation, the undetected 
 antineutrino is, in consequence, assumed to be moving in the opposite
 direction to the electron.  
 Altogether leads to the conclusion that the produced 
 electron has negative helicity and the antineutrino has positive helicity.
 Therefore, the charged weak currents responsible for these decays always
 produce left-handed electrons and right-handed antineutrinos. The
 non-observation of left-handed antineutrinos nor right-handed neutrinos in
 processes mediated by charged weak interactions is a 
 signal of parity violation since the parity
 transformation changes a left-handed fermion into the corresponding 
 right-handed fermion and viceversa. In fact, it is an indication of
 {\it maximal
 parity violation } which implies that the charged weak current must be
 neccessarily of the vector minus axial vector form,
 
 \begin{center}
 \framebox{$J_{\mu}\sim V_{\mu}-A_{\mu}$}
 \end{center}
 
 Let us see this in more detail. The vector and axial vector currents
 transform under parity as follows,
 
 $$V^{\mu}=\overline{\Psi}\gamma^{\mu}\Psi\stackrel{P}{\longrightarrow} 
 \left\{ \begin{array}{l}
 +\overline{\Psi}\gamma^0\Psi \\ -\overline{\Psi}\gamma^k\Psi \;;\; k=1,2,3
 \end{array} \right.$$
 \vspace{0.2cm}
 $$A^{\mu}=\overline{\Psi}\gamma^{\mu}\gamma^5\Psi\stackrel{P}{\longrightarrow} 
 \left\{ \begin{array}{l}
 -\overline{\Psi}\gamma^0\gamma^5\Psi \\
 +\overline{\Psi}\gamma^k\gamma^5\Psi \;;\; k=1,2,3
 \end{array} \right.$$
 Therefore the various products transform as,
 $$V_{\mu}V^{\mu}\stackrel{P}{\longrightarrow}V_{\mu}V^{\mu}$$  
 $$A_{\mu}A^{\mu}\stackrel{P}{\longrightarrow}A_{\mu}A^{\mu}$$   
 $$A_{\mu}V^{\mu}\stackrel{P}{\longrightarrow}-A_{\mu}V^{\mu}$$
  
 Any combination of vector and axial vector currents as $J_{\mu}\sim \alpha
 V_{\mu}+\beta A_{\mu}$ will generate parity violation in the Lagrangian,
 ${\cal L} \sim J_{\mu}J^{\mu +}$. But maximal parity violation is only reached
 if $J_{\mu}\sim V_{\mu}-A_{\mu}$, since 
 $$J_{\mu}J^{\mu +}\sim (V_{\mu}-A_{\mu})(V^{\mu}-A^{\mu})
 \stackrel{P}{\longrightarrow}(V_{\mu}+A_{\mu})(V^{\mu}+A^{\mu})$$
 which translates into that charged weak interactions only couple to left-handed
 fermions or right-handed antifermions. This can be seen simply
  by rewritting the
 current $J_{\mu}$ in terms of the field components. For instance, the leptonic
 current can be rewritten in terms of the left-handed fields as,
 $$J_{\mu}\sim V_{\mu}-A_{\mu}=\overline{\nu}_e\gamma_{\mu}(1-\gamma_5)e=
 2\overline{(\nu_e)_L}\gamma_{\mu}e_L$$ 
   
\subsection{V-A Theory of Charged Weak Interactions}
After the discovery of parity violation in weak interactions, Feynman and 
Gell-Mann in 1958 proposed the V-A Theory \cite{FeynGell}
which incorporated the success of
the Fermi Theory and solved the question of parity non-conservation by
postulating instead a V-A form for the charged weak current. The
current-current interactions are, like in the Fermi Theory, of local character,
being contracted at the same space-time point. The effective weak copling is,
as in the Fermi Theory, given by the Fermi constant, $G_F$.   

The Lagrangian of the V-A Theory
for the two first fermion generations is as follows,
  
\vspace{0.3cm}
\begin{center}
\framebox{
${\cal L}_{V-A}=-\frac{G_F}{\sqrt{2}} J_{\mu}^{CC}(x)J^{\mu CC+}(x)$ }\\
\vspace{0.3cm}
\framebox{
$J_{\mu}^{CC}=\overline{\nu}_e\gamma_{\mu}(1-\gamma_5)e+
\overline{\nu}_{\mu}\gamma_{\mu}(1-\gamma_5)\mu+
\overline{u}\gamma_{\mu}(1-\gamma_5)d'$  }
\end{center}
\vspace{0.3cm}
Notice that the $d$-quark field appearing in this Lagrangian, denoted by $d'$,
is the weak interactions $d$-quark eigenstate which is different than the 
$d$-quark mass eigenstate, denoted in these lectures by $d$. They are related 
by a rotation of the so-called Cabibbo angle $\theta_c$,
$$d'=cos\theta_c d+sin\theta_c s$$

The idea of the rotated d-quark states was proposed by Cabibbo in 1963 
\cite{Cabi} to 
account
for weak decays of 'strange' particles and, in particular, to explain the
suppression factor of the kaon decay rate  as compared to the pion decay rate
which experimentally was found to be about $\frac{1}{20}$. By comparing the
theoretical prediction from the V-A Theory with the experimental data, the
numerical value of the $\theta_c$ angle is inferred,

\vspace{0.3cm}
\begin{center}
$\frac{\Gamma(K^-\rightarrow \mu^-\overline{\nu}_{\mu})}
   {\Gamma(\pi^-\rightarrow \mu^-\overline{\nu}_{\mu})} 
   \sim \frac{sin^2\theta_c}{cos^2\theta_c}\sim\frac{1}{20}$
\hspace{0.5cm}$\Longrightarrow$\hspace{0.5cm}
\framebox{$\theta_c\simeq 13^{\circ}$}
\end{center}
\vspace{0.3cm}

The value of the effective coupling of the weak interactions, $G_F$ is deduced
from the meassurement of the $\mu$ lifetime,
 $$\tau_{\mu}^{\rm exp}=2.2\times 10^{-6}\;sec$$
 The prediction in the V-A Theory
to tree level and by neglecting the electron mass is,

$$\frac{1}{\tau_{\mu}}=\Gamma(\mu^-\rightarrow e^-\overline{\nu}_e\nu_{\mu})=
  \frac{G_F^2m_{\mu}^5}{192 \pi^3}$$
 and from this, 
\vspace{0.2cm}
\begin{center}
\framebox{$G_F=1.167\times 10^{-5}\; GeV^{-2}$}
\end{center}
\vspace{0.2cm}

The V-A Theory described reasonably well the phenomenology of weak interactions
until the discovery of the neutral currents in 1973 \cite{NC}. 
Notice that the neutral currents
were not included in the formulation of the V-A Theory. Besides, The V-A Theory
presented some non-appealing properties from the point of view of the
consistency of the theory itself. In particular, the V-A Theory violates
unitarity and it is a non-renormalizable theory. The unitarity violation
property can be seen, for instance, by comparing the prediction in the 
V-A Theory of the cross
section for elastic scattering of electron and neutrino,
$$\sigma_{V-A}(\nu e^- \rightarrow \nu e^-)= \frac{G_F^2}{6\pi} s$$
with the unitarity bound for the total cross section 
which is obtained from the general requirement of
unitarity of the scattering S-matrix,

$$SS^+=S^+S=I\hspace{0.2cm}\Longrightarrow\hspace{0.2cm}|a_J(s)|^2\leq 1\;
\forall J
\hspace{0.2cm}\Longrightarrow $$
$$\sigma(s)_{\rm tot}=\frac{16\pi}
{s}\sum_J(2J+1)|a_J(s)|^2\leq  \frac{16\pi}{s}\sum_J(2J+1)$$

It is clear that for high energies the prediction from the V-A Theory 
surpasses the
unitarity bound and, therefore, it should not be trusted.
It happens roughly at $\sqrt{s}\sim 300\;GeV$. 

The non-renormalizability of the
V-A Theory can be seen, for instance,  by computing loop contributions to 
the cross section and realizing that there appear quadratic divergences which 
cannot be absorved into redefinitions of the parameters of this theory. As in
the previous discussion on unitarity, it is due to the 'bad' behaviour of the
V-A Theory at high energies. The V-A Theory is said to be non-predictive at
high energies and it should only be used as an effective theory at low enough 
energies.  

\subsection{Intermediate Vector Boson Theory}
The Intermediate Vector Boson (IVB) Theory of weak interactions  assumed that
these are mediated by the exchange of  massive vector bosons with spin, $s=1$.
First, it was
 proposed the existence of intermediate charged vector bosons $W^{\pm}$ for
the charged weak interactions 
\cite{LeeYang2} and later the intermediate neutral vector boson $Z$ for the 
 neutral weak interactions \cite{Glash}. 
Notice that these bosons were not true gauge bosons yet.

The interaction Lagrangian of the IVB Theory, including both the charged (CC) and the 
neutral (NC) currents, is given by,

\begin{center} 
\framebox{
\shortstack{
${\cal L}_{IVB}={\cal L}_{CC}+{\cal L}_{NC}$\\
${\cal L}_{CC}=\frac{g}{\sqrt{2}}(J_{\mu}W^{\mu+}+J_{\mu}^{+}W^{-\mu})$\\
${\cal L}_{NC}=\frac{g}{cos\theta_w}J_{\mu}^{NC}Z^{\mu}$
   }
     } 
\end{center} 

where,
 
\begin{center}
\vspace{0.2cm}
\framebox{
$J_{\mu}=\sum_l \overline{\nu}_l\gamma_{\mu}
\left(\frac{1-\gamma_5}{2}\right)l+
\sum_q\overline{q}\gamma_{\mu}\left(\frac{1-\gamma_5}{2}\right)q'$ }
\vspace{0.2cm}
\framebox{
$J_{\mu}^{NC}=\sum_{f=l,q} g_L^f \overline {f}\gamma_{\mu}
\left(\frac{1-\gamma_5}{2}\right)f+
\sum_{f\neq\nu}g_R^f\overline{f}\gamma_{\mu}
\left(\frac{1+\gamma_5}{2}\right)f $}
\vspace{0.2cm}
\end{center}  
  
Here the $W^{\pm}_{\mu}$ and $Z_{\mu}$ are the charged and neutral 
intermediate vector bosons respectively and $g$ is the dimensionless
weak coupling.  The weak angle, $\theta_w$, defines the rotation 
in the neutral sector
from the weak eigenstates to 
the physical mass eigenstates,
 and relates the weak coupling to the electromagnetic coupling,
$g=\frac{e}{sin\theta_w}$.
 
Notice that the current-current
interactions are non-local, in contrast to the V-A Theory, due to the
propagation of the intermediate bosons. Besides, the new proposed neutral
currents have both V-A and V+A components, although experimentaly it is known
that the V-A component dominates. 

The prediction of neutral currents in 1961 \cite{Glash}
was corroborated experimentally 12
years later! in neutrino-hadron scattering by the Gargamelle collaboration at
CERN \cite{NC}. 
It was a great success of the IVB Theory which was incorporated later
into the construction of the SM. 

The relation between the parameters of the IVB Theory and the V-A Theory, which
is also incorporated in the construction of the SM, can be
 obtained by
comparison of the predictions from the two theories for $e\nu\rightarrow e\nu$
scattering at low energies ($\sqrt{s}<<M_W$),

\begin{center}
\vspace{0.2cm}
\framebox{
$\frac{G_F}{\sqrt{2}}=\frac{g^2}{8M_W^2}$  }
\vspace{0.2cm}
\end{center}
   
Finally, the IVB Theory is not free of problems either. It shares with the V-A
Theory the problems of non-renormalizability and violation of unitarity at high
energies. At low energies, say below the $M_W$ threshold, the IVB Theory is 
a well behaved effective theory of the weak interactions, but above it the
theory behaves badly. The problem of non-renormalizability can be seen for
instance by studing the $e^+e^-\rightarrow e^+e^-$ scattering proccess at one
loop. There are one-loop diagramms with $W$ bosons propagating in the internal
lines that diverge quadratically at high energies due to the bad behaviour of
the $W$ boson propagator in the IVB Theory,
$$\left( -i\Delta_W \right)_{IVB}\stackrel{k^2>>M_W^2}{\longrightarrow} 
\frac{1}{M_W^2} $$ 
This should be compared with the well behaved $W$ boson propagator  
in the Electroweak Gauge Theory,
$$\left( -i\Delta_W\right)_{\rm gauge}\stackrel{k^2>>M_W^2}{\longrightarrow} 
\frac{1}{k^2} $$ 
The violation of the unitarity bound occurs at slightly higher energies that in
the V-A Theory case. For instance, the cross-section for the production of
two longitudinal gauge bosons from neutrinos in the IVB Theory at tree level 
is,
$$\sigma_{IVB}(\nu\overline{\nu}\rightarrow W_L^+W_L^-)\sim 
 \frac{g^4}{M_W^4}s$$
which surpasses the unitarity bound at approximately, $\sqrt{s}\sim 500\;GeV$.
Notice that there is just one contributing diagramm, the one
with an electron in the t-channel.  Notice also that the IVB Theory does not include
the vector bosons self-interactions which are generic of non-abelian gauge
theories. These are precisely the 'repairing' interactions ocurring in the 
Electroweak Theory . The prediction from the SM for the previous
scattering proccess includes the contribution from an extra diagramm with a $Z$
boson exchanged in the s-channel which couples to the final $W_L^+W_L^-$ pair
with a typical non-abelian Yang Mills coupling. This new diagramm cancels the
bad high energy behaviour of the previous one. This dramatic cancellation also
occurs in many other proccesses. See, for instance, the meassurement of the
cross-section for $e^+e^-\rightarrow W^+W^-$ at LEP presented in Nodulman's
lectures, where these cancellations are shown.

\section{Building The Electroweak Theory}
\subsection{Some notes on History}
The proposal of the symmetry group for the Electroweak Theory,
$\ws$, was done by Glashow in 1961 \cite{Glash}. 
His motivation was rather to unify weak and
electromagnetic interactions into a symmetry group that contained $\em$. The
predictions included the existence of four physical vector boson eigenstates, 
$W^{\pm}$, $Z$, and $\gamma$,
obtained from rotations of the weak eigenstates. In particular, the rotation by
the weak angle $\theta_w$ which defines the $Z$ weak boson was introduced
already in this work. The massive weak bosons $W^{\pm}$ and $Z$ were 
considered as the exchanged  bosons
in the weak interactions, but they were not considered yet as gauge bosons. The
vector boson masses $M_W$ and $M_Z$ were parameters introduced by hand and
the interaction Lagrangian was that of the IVB Theory. 

Another key ingredient for the building of the Electroweak Theory is provided
by the
Goldstone Theorem which was initiated by Nambu in 1960 and proved and
studied with generality  
by Goldstone in 1961 and by Goldstone, Salam and Weinberg in 1962 \cite{gold}. 
This theorem states the existence of massless spinless
particles as an implication of spontaneous symmetry breaking of global
symmetries. 
        
The spontaneous symmetry breaking of local (gauge) symmetries, needed for
the breaking of the electroweak symmetry $\ws$, was studied by P. Higgs,
F.Englert and R.Brout, Guralnik, Hagen and Kibble in 1964 and later 
\cite{higgs}. 
These works were
inspired in previous studies within the context of condensed-matter
 physics as those
by Nambu and Jona-Lasinio on BCS Theory of superconductivity and works by
Schwinger in 1962 and by Anderson in 1963 \cite{condmatt}. 
The procedure for this spontaneous 
breakdown of gauge symmetries is referred to as the Higgs Mechanism. 

The Electroweak Theory as it is known nowadays was formulated by Weinberg in
1967 and by Salam in 1968 who incorporated the idea of unification of Glashow
\cite{gws}. This Theory, commonly called 
Glashow-Weinberg-Salam Model or 
SM, was
built with the help of the gauge principle and the knowledgde of gauge theories
and incorporated all the good phenomelogical properties of the pregauge
theories of the weak interactions, and in particular those of the IVB theory.
 The SM is indeed
a gauge theory based on the gauge symmetry of the electroweak interactions
$\ws$ and the intermediate vector bosons, $\gamma$, $W^{\pm}$ and $Z$ are the
four associated gauge bosons. The gauge boson masses, $M_W$ and $M_Z$, are
generated by the Higgs Mechanism in the Electroweak Theory and, as a 
consequence, it respects unitarity
at all energies and is renormalizable.

The important proof of renormalizability of gauge theories with and without
spontaneous symmetry breaking was provided by 't Hooft in 1971 \cite{hooft}.     

The first firm indication that the Sandard Model was the correct theory of
electroweak interactions was probably the  discovery of Neutral Currents in
1973 \cite{NC}
which included the first meassurement of $sin^2\theta_w$. By using this
experimental input for $\theta_w$ and the values of the electromagnetic 
coupling and $G_F$, the SM provided the first estimates for $M_W$ and $M_Z$
at that time which were already very close to the present values. 

Another important ingredients of the SM are: fermion family replication, quark
mixing and CP violation. After the proposal of the $d-s$ quark mixing given by
the Cabibbo angle \cite{Cabi}, the charm quark was postulated \cite{BjGlas} as
the companion of the $s$ quark in the charged weak interactions. Futhermore,
Glashow, Iliopoulos and Maiani showed in 1970 \cite{GIM} that any sensible weak
interaction theory must have this extra associated hadronic current in order 
to suppress to an acceptable level the induced strangeness-changing-neutral
current effects. This suppression mechanism of flavour-changing-neutral
currents (FCNC), usualy called GIM Mechanism, although invented before the
general acceptance of gauge theories, can best be explained in that context.
The existence of the $c$ quark was confirmed in 1974 \cite{charm} with the
discovery of the $J-\Psi$ particle which is interpreted as a $c\overline{c}$
bound state. With the discovery of the $\tau$ and $\nu_{\tau}$ leptons 
\cite{tau} and the $b$ quark \cite{b}, the fermion scenario with three families
was set in. Finally, the discovery of the top quark in 1994 (17 years later!)
\cite{top} has completed this scenario. The quark mixing in the three
generations case is given by the so-called Cabibbo-Kobayashi-Maskawa matrix
\cite{CKM} which incorporates the needed phase for CP violation in the SM.

The gold success of the SM was clearly the discovery of the gauge
bosons $W^{\pm}$ and $Z$ at the SpS collider at CERN in 1983 \cite{SpS}. 
Since then there
have been plenty of succesfull tests of the SM.

\subsection{Choice of the group $\ws$}
In order to follow the argument for the choice of the relevant group in the
Electroweak Theory, $\ws$, it is sufficient to consider the $e^-\nu_e$ component
of the charged weak current that we write now in the form, 
$$J_{\mu}=\overline{\nu}\gamma_{\mu}\left(\frac{1-\gamma_5}{2}\right)e=
\overline{\nu_L}\gamma_{\mu}e_L=\overline{l_L}\gamma_{\mu}\sigma_+l_L$$
$$J_{\mu}^+=\overline{e}\gamma_{\mu}\left(\frac{1-\gamma_5}{2}\right)\nu=
\overline{e_L}\gamma_{\mu}\nu_L=\overline{l_L}\gamma_{\mu}\sigma_-l_L$$
and we have introduced the lepton doublet notation and the
$\sigma_i\;(i=1,2,3)$ Pauli matrices,
$$l_L=\left( \begin{array}{c} \nu_L \\ e_L \end{array}\right)\;,\;
\overline{l_L}=\left(\begin{array}{cc}
\overline{\nu_L}&\overline{e_L}\end{array} \right)\;,\;
\sigma_{\pm}=\frac{1}{2}(\sigma_1\pm i\sigma_2)$$
$$\sigma_1=\left(\begin{array}{cc}0&1\\1&0\end{array}\right)\;,\;
\sigma_2=\left(\begin{array}{cc}0&-i\\i&0\end{array}\right)\;,\;
\sigma_3=\left(\begin{array}{cc}1&0\\0&-1\end{array}\right)$$
 
The $2\times 2$ matrices $T_i=\frac{\sigma_i}{2}\; i=1,2,3$ are the three
generators of $SU(2)$. 

Notice that in the charged currents there are just two generators $T_1$ and $T_2$.
A third generator $T_3$ is needed in order to close the $SU(2)$ algebra.
This implies the formulation of the third current that is relevant for
electroweak interactions,
$$J_{\mu}^3=\overline{l_L}\gamma_{\mu}\frac{\sigma_3}{2}l_L=
\frac{1}{2}(\overline{\nu_L}\gamma_{\mu}\nu_L-\overline{e_L}\gamma_{\mu}e_L)$$

The weak isospin group is the $SU(2)$ group that is generated by these three
generators and is usually denoted by $SU(2)_L$, where the subscript $L$ refers
to the left-handed character of the three weak currents. The weak isospin
algebra is correspondingly,
$$\left[\frac{\sigma_i}{2},\frac{\sigma_j}{2}\right]=i\epsilon_{ijk}
\frac{\sigma_k}{2}$$
where the $SU(2)$ structure constants are the completely antisymmetric
Levi-Civita symbols $\epsilon_{ijk}$. 

By Noether's Theorem 
 there are three associated conserved weak charges,
$$T^i=\int d^3xJ_0^i(x)\;,\;i=1,2,3$$    

It is interesting to notice that the above introduced neutral weak current
is none of the two physical known neutral currents, $J_{\mu}^{em}$ and 
$J_{\mu}^{NC}$. Futhermore, none of these two currents have definite properties
under $SU(2)_L$ transformations, whereas $J_{\mu}^3$ does. With the motivation
of unifying the electromagnetic and weak interactions, Glashow proposed to
include the electromagnetic current by adding to $SU(2)_L$ 
a new $U(1)$ group which should
be different than $\em$ in order to the get the proper conmutation relations
among the $U(1)$ and $SU(2)_L$ generators. The new proposed group is the 
weak hypercharge group $U(1)_Y$ with one generator $\frac{Y}{2}$ which indeed,
 as it must be, conmutes with the three $SU(2)_L$ generators. The associated
 neutral current is the weak hypercharge current, $J_{\mu}^Y$, and the conserved
  charge is the weak hypercharge $Y$. 
Within this formalism there is some sort of
electromagnetic and weak interactions unification  
since the $\em$ group appears
as a subgroup of the total electroweak group,
$$U(1)_{em} \subset \ws$$
The relation among the charges associated to the three neutral currents,
$J_{\mu}^{em}$, $J_{\mu}^3$ and $J_{\mu}^Y$, is a replica of the Gell-Mann
Nishijima relation,
\vspace{0.2cm}  
\begin{center} 
\framebox{$Q=T_3+\frac{Y}{2}$}
\end{center} 
\vspace{0.2cm}
where now, 
$$Q={\rm electric\; charge}\;,\;T_3={\rm weak\; isospin}\;,\;Y={\rm weak\;
hypercharge};$$
and the corresponding relation among the currents is,
$$J_{\mu}^{em}=J_{\mu}^3+J_{\mu}^Y$$
Therefore, if the following are used as inputs 
$$J_{\mu}^{em}=(-1)\overline{e_L}\gamma_{\mu}e_L
      +(-1)\overline{e_R}\gamma_{\mu}e_R$$
$$J_{\mu}^3=\left(-\frac{1}{2}\right)\overline{e_L}\gamma_{\mu}e_L
   +\left(\frac{1}{2}\right)\overline{\nu_L}\gamma_{\mu}\nu_L $$
one can get $J_{\mu}^Y$ and the orthogonal combination $J_{\mu}^{NC}$ as 
outputs,
$$J_{\mu}^Y=2(J_{\mu}^{em}-J_{\mu}^3)=(-1)\overline{e_L}\gamma_{\mu}e_L+
 (-2)\overline{e_R}\gamma_{\mu}e_R+(-1)\overline{\nu_L}\gamma_{\mu}\nu_L$$
 $$J_{\mu}^{NC}\perp J_{\mu}^{em}\Rightarrow $$
$$ J_{\mu}^{NC}=c_w^2 J_{\mu}^3-s_w^2 \frac{J_{\mu}^Y}{2}=
   (-\frac{1}{2}+s_w^2)\overline{e_L}\gamma_{\mu}e_L+
   (s_w^2)\overline{e_R}\gamma_{\mu}e_R+
   (\frac{1}{2})\overline{\nu_L}\gamma_{\mu}\nu_L $$
where,
$$c_w=cos\theta_w\;;\; s_w=sin\theta_w\;;\;\theta_w={\rm weak\; angle}$$
Notice that the currents that couple to the physical neutral bosons 
$A_{\mu}$ and $Z_{\mu}$ are $J_{\mu}^{em}$ and $J_{\mu}^{NC}$ respectively,
and it is this last one, $J_{\mu}^{NC}$, that inherits the generic name 
of neutral current.

From the above expressions for the neutral currents, one can also extract the
values of the corresponding charges and couplings. For instance, from
$J_{\mu}^{NC}$ one gets the relevant factors in the weak neutral couplings to
electrons and neutrinos , $g_L^e=-\frac{1}{2}+s_w^2$, $g_R^e=s_w^2$, 
$g_L^{\nu}=\frac{1}{2}$ and $g_R^{\nu}=0$. 

If one includes the contributions from all the quarks and leptons of the three
families, the neutral currents are written generically as:

$$J_{\mu}^Y=\sum_f Y_{f_L}\overline{f_L}\gamma_{\mu}f_L+
     \sum_f Y_{f_R}\overline{f_R}\gamma_{\mu}f_R$$
$$J_{\mu}^{NC}=\sum_f g_L^f \overline{f_L}\gamma_{\mu}f_L+
     \sum_{f\neq \nu} g_R^f \overline{f_R}\gamma_{\mu}f_R$$
$$J_{\mu}^{em}=\sum_f Q_f \overline{f_L}\gamma_{\mu}f_L+     
     \sum_f Q_f \overline{f_R}\gamma_{\mu}f_R$$
$$J_{\mu}^3=\sum_f T_3^f \overline{f_L}\gamma_{\mu}f_L $$

where,
$$g_L^f=T_3^f-Q_fs_w^2\;\;;\;\;g_R^f=-Q_fs_w^2$$
 The corresponding quantum numbers for the fermions of the first family are
collected in Tables 1 and 2. The fermions of the second and third family have 
the same quantum numbers as the corresponding fermions of the first one. 
\begin{table}[htb]
\begin{center}
\caption{Lepton quantum numbers }
\begin{tabular}{crrrr}
\hline
Lepton & $T$ & $T_3$ & $Q$ & $Y$\\
\hline
$\nu_L$& $\frac{1}{2}$ & $\frac{1}{2}$ & $0$ & $-1$ \\
$e_L$ & $\frac{1}{2}$ & $-\frac{1}{2}$ & $-1$ & $-1$\\
$e_R$ & $0$ & $0$ & $-1$ & $-2$ \\
\hline
\end{tabular}
\end{center}
\end{table}
\begin{table}[htb]
\begin{center}
\caption{Quark quantum numbers }
\begin{tabular}{crrrr}
\hline
Quark & $T$ & $T_3$ & $Q$ & $Y$\\
\hline
$u_L$& $\frac{1}{2}$ & $\frac{1}{2}$ & $\frac{2}{3}$
&$\frac{1}{3}$  \\
$d_L$ & $\frac{1}{2}$ & $-\frac{1}{2}$ & $-\frac{1}{3}$ &
$\frac{1}{3}$\\
$u_R$ & $0$ & $0$ & $\frac{2}{3}$ & $\frac{4}{3}$ \\
$d_R$ & $0$ & $0$ & $-\frac{1}{3}$ & $-\frac{2}{3}$\\
\hline
\end{tabular}
\end{center}
\end{table}

Similarly, the charged current is written generically as,
$$J_{\mu}=\sum_f\overline{f_L}\gamma_{\mu}\sigma_+f_L$$
  
Finally, the electroweak interaction Lagrangian is  written in
 terms of the currents 
and the physical fields as,
$${\cal L}_{\rm int}= {\cal L}_{CC}+{\cal L}_{NC}+{\cal L}_{em}$$
where,  
$${\cal L}_{CC}=\frac{g}{\sqrt{2}}(J_{\mu}W^{\mu\;+}+J_{\mu}^+W^{\mu\;-})$$
$${\cal L}_{NC}=\frac{g}{c_w}J_{\mu}^{NC}Z^{\mu}$$
$${\cal L}_{em}=eJ_{\mu}^{em}A^{\mu}$$
Notice that ${\cal L}_{CC}$ and ${\cal L}_{NC}$ are the same Lagrangians as in 
the IVB Theory.

\section{SM: The Gauge Theory of Electroweak Interactions}

 The SM is the gauge theory for electroweak interactions and
 has provided plenty of successful predictions with an impressive level of
 precision.
 
 It is based on the gauge symmetry of electroweak interactions, namely, the
 symmetry $\ws$ previously introduced which is required to be a local
 symmetry of the electroweak Lagrangian. As before, $SU(2)_L$ is the weak
 isospin group which acts just on left-handed fermions and $U(1)_Y$ is the
 weak hypercharge group. The $\ws$ group has four generators, three of
 which are the $SU(2)_L$ generators, $T_i=\frac{\sigma_i}{2}$ with
 $i=1,2,3$, and the fourth one is the $U(1)_Y$ generator, $\frac{Y}{2}$. The
 conmutation relations for the total group are:
 $$\left[T_i,T_j\right]=i\epsilon_{ijk}T_k\;\;;\;\;
  \left[T_i,Y\right]=0\;\;;\;\; i,j,k=1,2,3 $$ 
 The left-handed fermions transform as doublets under $SU(2)_L$,
 $$f_L\rightarrow e^{i\vec{T}\vec{\theta}}f_L\;\;;\;\;
 f_L=\left(\begin{array}{c}\nu_L\\e_L\end{array}\right)\;,\;
 \left(\begin{array}{c}u_L\\d_L\end{array}\right),...$$
 whereas the right-handed fermions transform as singlets,
 $$f_R\rightarrow f_R\;;\;f_R=e_R\;,\;u_R\;,\;d_R,...$$
 The fermion quantum numbers are as in Tables 1 and 2 and the relation
 $$Q=T_3+\frac{Y}{2}$$
 is also incorporated in the SM. 
 
 The number of associated gauge bosons, being equal to the number of
 generators, is four:
  
 $\underline{{\rm gauge\; bosons}}$\\
 \begin{center}
 $W_{\mu}^i\;,\;i=1,2,3$. These are the weak bosons of $SU(2)_L$\\
 $B_{\mu}$. This is the hypercharge boson of $U(1)_Y$
 \end{center}
       
 The building of the SM Lagrangian is done by following the same steps as
 in any gauge theory. In particular, the $\ws$ symmetry is promoted from
 global to local by replacing the derivatives of the fields by the
corresponding covariant derivatives. For a generic fermion field $f$, its
covariant derivative corresponding to the $\ws$ gauge symmetry is,
$$D_{\mu}f=\left(\partial_{\mu}-ig\vec{T}.\vec{W}_{\mu}-ig'\frac{Y}{2}
B_{\mu}\right)f$$
where,
\begin{center}
$g$ = coupling constant corresponding to $SU(2)_L$\\
$g'$= coupling constant corresponding to $U(1)_Y$
\end{center}     
For example, the covariant derivatives for a left-handed and a right-handed
electron are respectively,
$$D_{\mu}e_L=
\left(\partial_{\mu}-ig\frac{\vec{\sigma}}{2}.\vec{W}_{\mu}+ig'\frac{1}{2}B_{\mu}
\right)e_L\;;\;D_{\mu}e_R=\left(\partial_{\mu}+ig'B_{\mu}\right)e_R$$
As in the previous cases of QED and QCD, the gauge invariant electroweak 
interactions are generated from the $\overline{f}i D{\hspace{-6pt}\slash}f$
term. After replacing the  covariant derivative above, and by
rotating the weak bosons to the physical basis, one can check that the  
interaction terms obtained for the electroweak bosons with the quarks and
leptons are the same as those in the interaction Lagrangian 
${\cal L}_{\rm int}$ 
given in the previous section.

\section{Lagrangian of The Electroweak Theory I}

In order to get the total Lagrangian of the Electroweak Theory one must add to
the previous fermion terms containing the kinetic and fermion interaction terms,
the gauge boson kinetic terms and the gauge boson
self-interaction terms. The SM total Lagrangian can be written as,

\begin{center}
\framebox{
${\cal L}_{SM}={\cal L}_f+{\cal L}_{G}+{\cal L}_{SBS}+{\cal L}_{YW}$}
\end{center}
where, the fermion Lagrangian is,

\begin{center}
\framebox{
${\cal L}_f=\sum_{f=l,q}\overline{f}i D{\hspace{-6pt}\slash}f$  }
\end{center}
and the Lagrangian for the gauge fields is,

\begin{center}

\framebox{
${\cal L}_{G}=-\frac{1}{4}W_{\mu\nu}^iW^{\mu\nu}_i-
\frac{1}{4}B_{\mu\nu}B^{\mu\nu}+{\cal L}_{GF}+{\cal L}_{FP}$ }
\end{center}
which is written in terms of the field strength tensors,
$$W_{\mu\nu}^i=\partial_{\mu}W_{\nu}^i-\partial_{\nu}W_{\mu}^i+
g\epsilon^{ijk}W_{\mu}^jW_{\nu}^k$$
$$B_{\mu\nu}=\partial_{\mu}B_{\nu}-\partial_{\nu}B_{\mu}$$
${\cal L}_{GF}$ and ${\cal L}_{FP}$ are the gauge fixing and Faddeev Popov 
Lagrangians respectively that are needed in any gauge theory. We omit to write
them here for brevity. These have also been omited in the cases of
QCD and QED.

Notice that this  gauge Lagrangian contains the wanted self-interaction terms
among the three $W_{\mu}^i\;,i=1,2,3$ gauge bosons, as it corresponds to a
non-abelian $SU(2)_L$ group. 

The last two terms, ${\cal L}_{SBS}$ and ${\cal L}_{YW}$ are the Symmetry
Breaking Sector Lagrangian and the Yukawa Lagrangian respectively. As will be
discussed in the forthcomming sections, these terms are needed in order to
provide  
the wanted $M_W$ and $M_Z$ gauge boson masses and $m_f$ fermion masses. 
      
One can show that ${\cal L}_{SM}$ is indeed invariant under the following 
$\ws$ gauge transformations:
$$\begin{array}{lll}
  f_L & \rightarrow & e^{i\vec{T}\vec{\theta}(x)}f_L \\
  f_R & \rightarrow & f_R \\
  f   & \rightarrow & e^{i\frac{Y}{2}\alpha(x)}f\\
  W_{\mu}^{i}& \rightarrow & W_{\mu}^{i}-
  \frac{1}{g}\partial_{\mu}\theta^i(x)+\epsilon^{ijk}\theta^jW_{\mu}^k\\
  B_{\mu} & \rightarrow &B_{\mu}-\frac{1}{g'}\partial_{\mu}\alpha(x)
  \end{array}$$
The physical gauge bosons $W_{\mu}^{\pm}$, $Z_{\mu}$ and $A_{\mu}$
 are obtained from the electroweak interaction 
eigenstates by the following expressions,

\begin{center}
\framebox{
\shortstack{
$W_{\mu}^{\pm}=\frac{1}{\sqrt{2}}(W_{\mu}^1\mp iW_{\mu}^2)$ \\
$Z_{\mu}=c_wW_{\mu}^3-s_wB_{\mu}$ \\
$A_{\mu}=s_wW_{\mu}^3+c_wB_{\mu}$ 
   }
     } 
\end{center}

where, $\theta_w$ defines the rotation in the neutral sector.  The relations
among the various couplings are obtained by identifying the interactions
terms with those of ${\cal L}_{\rm int}$. Thus one gets,

\begin{center}
\framebox{
$g=\frac{e}{s_w}$} \hspace{0.2cm}\framebox{ $g'=\frac{e}{c_w}$ }
\end{center}
 
Finally, note that mass terms as $M_W^2W_{\mu}W^{\mu}$, 
$\frac{1}{2}M_Z^2Z_{\mu}Z^{\mu}$ and $m_f\overline{f}f$ are forbidden by 
$\ws$ gauge invariance. This is a new situation which is not found in QED or
QCD. The needed gauge boson masses must be generated in a gauge invariant
way. The spontaneous breaking of the $\ws$ symmetry and the 
Higgs Mechanism provide indeed this mass generation.  To this subject we come
next.

\section{The Concept of Spontaneous Symmetry Breaking and The Higgs Mechanism}

One of the key ingredients of the SM of electroweak
interactions  is the concept of Spontaneous Symmetry
Breaking (SSB), giving rise to Goldstone-excitations \cite{gold} which
in turn can be
related to gauge boson mass terms \cite{sbs}. When this SSB refers to   
a gauge symmetry instead of a global symmetry, then the  Higgs
Mechanism operates \cite{higgs}. This procedure is needed in order 
to describe the short
ranged weak interactions by a gauge theory without spoiling gauge
invariance. The discovery of the $W^{\pm}$ and $Z$ gauge bosons at CERN
in 1983 \cite{SpS}
may be considered as the first experimental evidence of the
SSB phenomenon in electroweak interactions. 
In present and future experiments one  hopes to get insight into the
nature of this Symmetry Breaking Sector (SBS) and this is one of the
main
motivations for constructing the next generation of accelerators. In
particular, it is the most exiciting challenge for the  
LHC collider being built at CERN.

In the SM, the symmetry breaking is realized linearly by a scalar field
which acquires a non-zero vacuum expectation value. The resulting
physical spectrum contains not only the massive intermediate vector
bosons and fermionic matter fields but also the Higgs particle, a
neutral scalar field which has escaped experimental
detection until now. The main advantage of the SM picture of
symmetry breaking lies in the fact that an explicit and consistent
formulation exists, and any observable can be calculated perturbatively
in the Higgs self-coupling constant.  However, the fact that one can
compute in a model doesn't mean at all that this is the right one.

The concept of spontaneous electroweak symmetry breaking is more general
than the way it is usually implemented in the SM. Any alternative
 SBS has a chance to replace the standard Higgs sector, provided it
meets the following basic requirements: 1) Electromagnetism
remains unbroken; 2) The full symmetry contains the electroweak gauge
symmetry; 3) The symmetry breaking occurs at about the energy scale
$v=(\sqrt{2}G_F)^{-\frac{1}{2}}=246\;GeV$ with $G_F$ being the Fermi
coupling constant.

In the following it is reviewed the basic ingredients of the symmetry breaking
phenomenom in the Electroweak Theory. Some relevant topics related with this
breaking are also discussed.
  
\subsection{The Phenomenon of Spontaneous Symmetry Breaking}
A simple definition of the phenomenon of SSB 
is as follows:

{\it A physical system has a symmetry that is spontaneously
broken if the interactions governing the dynamics of the system possess
such a symmetry but the ground state of this system does not}.

An illustrative example of this phenomenon is the infinitely extended
ferromagnet. For this purpouse, let us consider the system near the
Curie temperature $T_C$. It is described by an infinite set of
elementary spins whose interactions are rotationally invariant, but its
ground state presents two different situations depending on the value of
the temperature $T$.

\vspace{0.5cm}
{\bf Situation I}: $T>T_C$

\vspace{0.5cm}
The spins of the system are randomly oriented and as a consequence the
average magnetization vanishes: $\vec{M}_{\rm average}=0$. The ground
state with these disoriented spins is clearly rotationally invariant.

\vspace{0.5cm}
{\bf Situation II}: $T<T_C$

\vspace{0.5cm}
The spins of the system are all oriented parallely to some particular
but
arbitrary direction and the average magnetization gets a non-zero value:
$\vec{M}_{\rm average} \neq 0$ ({\it Spontaneous Magnetization}). Since
the
directions of the spins are arbitrary, there are infinite possible ground
states, each one corresponding to one possible direction and all having
the same (minimal) energy. Futhermore, none of these states are
rotationally invariant since there is a privileged direction. This is,
therefore, a clear example of SSB since the
interactions among the spins are rotationally invariant but the ground
state is not. More specifically, it is the fact that the system
'chooses' one among the infinite possible non-invariant ground states
what produces the phenomenon of SSB.

On the theoretical side, and irrespectively of what could be the origen
of such a physical phenomenon at a more fundamental level, one can
parametrize this behaviour by means of a symple mathematical model. In
the case of the infinitely extended ferromagnet one of these models is
provided by the Theory of Ginzburg and Landau \cite{gl}. We present in the
following the basic ingredients of this model.

For $T$ near $T_C$, $\vec{M}$ is small and the free energy density
$u(\vec{M})$ can be approached by (here higher powers of $\vec{M}$ are
neglected):
\begin{eqnarray}
u(\vec{M})& = &(\partial_i \vec{M})(\partial_i \vec{M})+V(\vec{M})\;;\;
i=1,2,3 \nonumber\\
V(\vec{M})&=&\alpha_1(T-T_C)(\vec{M}.\vec{M})+\alpha_2(\vec{M}.\vec{M})^2
\;;\;\alpha_1,\alpha_2>0\nonumber
\label{landau}
\end{eqnarray}
The magnetization of the ground state is obtained from the condition of
extremum:
\begin{eqnarray}
& &\frac{\delta V(\vec{M})}{\delta M_i}=0\Rightarrow 
\vec{M}.\left [ \alpha_1(T-T_C)+2\alpha_2(\vec{M}.\vec{M})\right ]=0
\nonumber
\label{min}
\end{eqnarray}
\begin{figure}
\vspace{1cm}
\hspace{1cm}\epsfysize7cm\epsffile{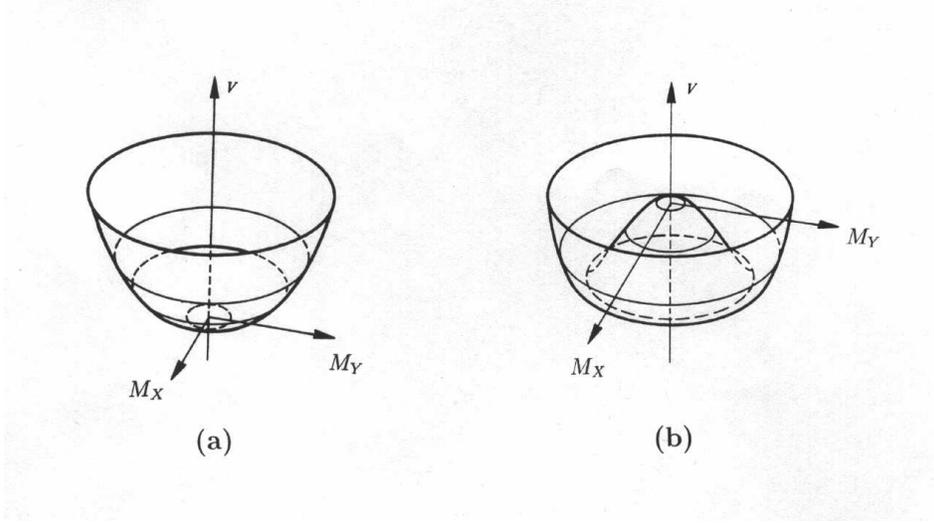}
\caption{The potential $V(\vec{M})$ in the symmetric phase (a)
 and in the spontaneously broken phase (b)}
\end{figure} 

There are two solutions for $\vec{M}$, depending on the value of $T$:

\vspace{0.5cm}
{\bf Solution I}:

\vspace{0.5cm}
If $T>T_C\; \Rightarrow \left [
\alpha_1(T-T_C)+2\alpha_2(\vec{M}.
\vec{M}) \right ]>0\;\Rightarrow \vec{M} =0$\\

The solution for $\vec{M}$ is the trivial one and corresponds to the
situation I described before where the ground state is rotational
invariant. The potential $V(\vec{M})$ has a symmetric shape with a
unique minimum at the origen $\vec{M}=0$ where $V(0)=0$. This is
represented in Fig.2a for the simplified bidimensional case,
$\vec{M}=(M_X,M_Y)$. 
 
\vspace{0.5cm}
{\bf Solution II}:

\vspace{0.5cm}
If $T<T_C \; \Rightarrow \vec{M}=0$ is a local maximum and
the condition of minimun requires:
$$ \alpha_1(T-T_C)+2 \alpha_2(\vec{M}.\vec{M})=0 \Rightarrow
|\vec{M}|=\sqrt{\frac{\alpha_1(T_C-T)}{2\alpha_2}}$$
Namely, there are infinite absolute minima having all the same
$|\vec{M}|$ above, but different direction of $\vec{M}$. This
corresponds to the situation II where the system has infinite possible
degenerate ground states which are not rotationally invariant. The
potential $V(\vec{M})$ has a 'mexican hat shape' as represented in
Fig.2b for the bidimensional case.
 
Notice that it is the choice of the particular ground state what
produces,
for $T<T_C$, the spontaneous breaking of the rotational symmetry.

\subsection{Spontaneous Symmetry Breaking  in Quantum Field Theory: QCD as
an example}
In the language of Quantum Field Theory, {\it a system is said to
possess
a symmetry that is spontaneously broken if the Lagrangian describing the
dynamics of the system is invariant under these symmetry
transformations, but the vacuum of the theory is not}. Here the vacuum
$|0>$ is the state where the Hamiltonian expectation value $<0|H|0>$ is
minimum.

For illustrative purposes we present in the following the particular
case of QCD where, besides the color gauge symmetry,  there is an extra 
global symmetry, named
chiral symmetry,  which turns out to be spontaneously broken . For
simplicity
let us consider QCD with just two flavours. The Lagrangian is that 
in section 5 with just the two ligthest quarks, $u$ and $d$.
 
One can check that for $m_{u,d}=0$, ${\cal L}_{QCD}$ has  
the chiral symmetry $\lr$ 
 that is defined by the following transformations:
\[ \Psi_L \rightarrow \Psi_L'=U_L \Psi_L \]
\[ \Psi_R \rightarrow \Psi_R'=U_R \Psi_R \]
where,
\[ \Psi = \left ( \begin{array}{c}u \\ d \end{array} \right ) \;;\;
\Psi_L=\frac{1}{2}(1-\gamma_5)\Psi\;;\;
\Psi_R=\frac{1}{2}(1+\gamma_5)\Psi \]
\[ U_L \in SU(2)_L \; ;\; U_R \in SU(2)_R \]

$U_L$ and $U_R$ can be written in terms of the 2x2 matrices $T_L^a$ and
$T_R^a$ ($a=1,2,3$) corresponding to the generators $Q_L^a$ and $Q_R^a$
of $SU(2)_L$ and $SU(2)_R$ respectively:
\[ U_L= \exp(-i\alpha_L^a T_L^a)\;;\;
   U_R= \exp(-i\alpha_R^a T_R^a)  \]

It turns out that the physical vacuum of QCD is not invariant under the
full chiral $\lr$ group but just under the subgroup $SU(2)_V= \lpr$ which
is precisely 
the already introduced  isospin group. The transformations given by the
axial subgroup, $SU(2)_A$, do not leave the QCD vacuum invariant.
Therefore, QCD with $m_{u,d}=0$ has a chiral symmetry which is
spontaneously broken down to the isospin symmetry:
\[ \lr \rightarrow SU(2)_V \]

The fact that in Nature $m_{u,d}\neq 0$ introduces an extra explicit
breaking of this chiral symmetry. Since the
fermion masses are small this explicit breaking is soft. The chiral
symmetry is not an exact but approximate symmetry of QCD.

One important question is still to be clarified. How do we know from
experiment that, in fact, the QCD vacuum is not $\lr$
symmetric?. Let us
assume for the moment that it is chiral invariant. We will see that this
assumption leads to a contradiction with experiment.

If $|0>$ is chiral invariant $\Rightarrow$
 \[  U_L|0>=|0>\;;\;U_R|0>=|0>  
\Rightarrow \]
\[ T_L^a|0>=0;T_R^a|0>=0 \Rightarrow Q_L^a|0>=0\;;\;
Q_R^a|0>=0 \]

In addition, if $|\Psi>$ is an eigenstate of the
Hamiltonian and parity operator such that:
\[ H|\Psi>=E|\Psi>\;;\;P|\Psi>=|\Psi> \]
then,
\[
\exists
|\Psi'>= \frac{1}{\sqrt{2}}(Q_R^a-Q_L^a)|\Psi>\; / \; H|\Psi'>=E|\Psi'>
\;;\;P|\Psi'>=-|\Psi'> \]
In summary, if the QCD vacuum is chiral invariant there must exist pairs
of degenerate states in the spectrum,
the so-called parity doublets as $|\Psi>$ and $|\Psi'>$,
which are related by a chiral transformation
and have opposite parities. The absence of such parity doublets in the
hadronic spectrum indicates that the chiral symmetry must be
spontaneously broken. Namely, there must exist some generators $Q^a$ of
the
chiral group such that $Q^a|0>\neq 0$. More specifically, it can be
shown that these generators are the three $Q^a_5 \; (a=1,2,3)$ of the
axial
group, $SU(2)_A$. In conclusion, the chiral symmetry breaking pattern in
QCD is $\lr \rightarrow SU(2)_V$ as announced.

\subsection{Goldstone Theorem}
One of the physical implications of the SSB
phenomenom is the appearance of massless modes. For instance, in the
case of the infinitely extended ferromagnet and below the Curie
temperature there appear modes connecting the different possible ground
states, the so-called spin waves.

The general situation in Quantum Fied Theory is described by the
Goldstone Theorem \cite{gold}:

{\it If a Theory has a global  symmetry of the Lagrangian which is not a
symmetry of the vacuum then there must exist one massless
boson, scalar or pseudoscalar, associated to each
generator which does not annihilate the vacuum and having its same
quantum numbers. These modes are referred to as Nambu-Goldstone bosons
or simply as Goldstone bosons.}

Let us return to the example of QCD. The breaking of the chiral symmetry
is characterized by $Q^a_5|0>\neq 0\; (a=1,2,3)$. Therefore, according
to Goldstone Theorem, there must exist three massless Goldstone
bosons, $\pi^a(x)\;\; a=1,2,3$, which are pseudoscalars. These bosons
are identified with the three physical pions.

The fact that pions have $m_{\pi}\neq 0$ is a consequence of the soft
explicit breaking in ${\cal L}_{QCD}$ given by $m_q \neq 0$. The fact that
$m_{\pi}$ is small and that there is a large gap between this mass and
the rest of the hadron masses can be seen as another manifestation of
the spontaneous chiral symmetry breaking with the pions being the
pseudo-Goldstone bosons of this breaking.

\subsection{Dynamical Symmetry Breaking}
In the previous sections we have seen the equivalence between the
condition $Q^a |0> \neq 0$ and the non-invariance of the vacuum under
the symmetry transformations generated by the $Q^a$ generators:
\[ U|0> \neq |0> \;;\; U= \exp(i\epsilon^a Q^a) \]

In Quantum Field Theory, it can be shown that an alternative way of
characterizing the phenomenom of SSB is by
certain field operators that have non-vanishing vacuum
expectation values (v.e.v.).
\[ {\rm SSB} \Longleftrightarrow \exists \Phi_j /
<0|\Phi_j|0> \neq 0 \]

This non-vanishing v.e.v. plays the role of the order parameter
signaling the existence of a phase where the symmetry of the vacuum is
broken.

There are several possibilities for the nature of this field operator.
In particular, when it is a composite operator which represents a
composite state being produced from a strong underlying dynamics, the
corresponding SSB is said to be a dynamical symmetry breaking. The
chiral symmetry breaking in QCD is one example of this type of breaking.
The non-vanishing chiral condensate made up of a quark and an anti-quark
is the order paremeter in this case:
\[ <0|\bar{q}q|0> \neq 0 \Rightarrow \lr \rightarrow SU(2)_V \]

The strong interactions of $SU(3)_C$ are the responsible for creating
these $\bar{q}q$ pairs from the vacuum and, therefore, the
$<0|\bar{q}q|0>$ should, in principle, be calculable from
QCD.

It is interesting to mention that this type of symmetry
breaking can happen similarly in more general $SU(N)$ gauge theories.
The corresponding gauge couplings become sufficiently strong at large
distances
and allow for spontaneous breaking of their additional chiral-like
symmetries. The corresponding order paremeter is also a chiral
condensate: $<0|\overline{\Psi}\Psi|0>\neq 0$.

\subsection{The Higgs Mechanism}
The Goldstone Theorem is for theories with spontaneously broken global
symmetries but does not hold for gauge theories. When a spontaneous
symmetry breaking takes place in a gauge theory the so-called Higgs
Mechanism operates \cite{higgs}:

{\it The would-be Goldstone bosons associated to the global symmetry
breaking do not manifest
explicitely in the physical spectrum but instead they 'combine' with the
massless gauge bosons and as result, once the spectrum of the theory is
built up on the asymmetrical vacuum, there appear massive vector
particles. The number of vector bosons that acquire a mass is
precisely equal to the number of these would-be-Goldstone bosons}.

There are three important properties of the Higgs Mechanism for
'mass generation' that are worth mentioning:
\begin{itemize}
\item[1.-]
It respects the gauge symmetry of the Lagrangian.
\item[2.-]
It preserves the total number of polarization degrees.
\item[3.-]
It does not spoil the good high energy properties nor the
renormalizability of the massless gauge theories.
\end{itemize}

We now turn to the case of the SM  of Electroweak
Interactions. We will see in the following how the Higgs
Mechanism
is implemented in the $\ws$ Gauge Theory in order to generate a mass for
the  weak gauge bosons, $W^{\pm}$ and $Z$.

The following facts must be
considered:
\begin{itemize}
\item[1.-]
The Lagrangian of the SM is gauge $\ws$ symmetric.
Therefore, anything we wish to add must preserve this symmetry.
\item[2.-]
We wish to generate masses for the three gauge bosons $W^{\pm}$ and $Z$
but not for the photon, $\gamma$. Therefore, we need three
would-be-Goldstone bosons, $\phi^+$, $\phi^-$ and $\chi$, which will
combine with the three massless gauge bosons of the $\ws$ symmetry.
\item[3.-]
Since $U(1)_{em}$ is a symmetry of the physical spectrum, it must be
a symmetry of the vacuum of the Electroweak Theory.
\end{itemize}

From the above considerations we conclude that in order to implement
the Higgs Mechanism in the Electroweak Theory we need to introduce
'ad hoc' an
additional system  that interacts with the gauge sector
in a $\ws$ gauge invariant manner and whose self-interactions, being
also introduced 'ad hoc', must produce the wanted breaking, $\ws
\rightarrow U(1)_{em}$, with the three associated would-be-Goldstone
bosons $\phi^+$, $\phi^-$ and $\chi$.
This sytem is the so-called SBS of the Electroweak Theory.


\section{The Symmetry Breaking Sector of the Electroweak Theory}
 
In this section we introduce and justify the simplest choice for the
SBS of the Electroweak Theory.

Let $\Phi$ be the additional system providing the $\ws \rightarrow
U(1)_{em}$ breaking. $\Phi$ must fulfil the following conditions:
\begin{itemize}
\item[1.-]
It must be a scalar field so that the above breaking preserves Lorentz
invariance.
\item[2.-]
It must be a complex field so that the Hamiltonian is hermitian.
\item[3.-]
It must have non-vanishing weak isospin and hypercharge in order to
break $SU(2)_{L}$ and $U(1)_{Y}$. The assignment of quantum
numbers and
the choice of representation of $\Phi$ can be done in many ways. Some
possibilities are:
\subitem -  Choice of a non-linear representation: $\Phi$
transforms non-linearly under $\ws$.
\subitem -  Choice of a linear representation: $\Phi$ transforms
linearly under $\ws$. The simplest linear representation is a complex
doublet. Alternative choices are: complex triplets, more than one
doublet, etc. In particular, one may choose two complex doublets $H_1$
and $H_2$ as in the Minimal Supersymmetric SM.
\item[4.-]
Only the neutral components of $\Phi$ are allowed to acquire a
non-vanishing v.e.v. in order to preserve the $U(1)_{em}$ symmetry
of the vacuum.
\item[5.-]
The interactions of $\Phi$ with the gauge and fermionic sectors
must be introduced in a gauge invariant way.
\item[6.-]
The self-interactions of $\Phi$ given by the potential $V(\Phi)$ must
produce the wanted breaking which is characterized in this case by
$<0|\Phi|0> \neq 0$. $\Phi$ can be, in principle, a fundamental or a
composite field.
\item[7.-]
If we want to be predictive from low energies up to very high energies the
interactions in $V(\Phi)$ must be renormalizable. 
\end{itemize}
By taking into account the above seven points one is led to the
following simplest choice for the system $\Phi$ and the Lagrangian of
the  SBS of the Electroweak Theory:
\begin{eqnarray}
\cl_{SBS}& =& (D_\mu \Phi)^\dagger (D^\mu \Phi) -V(\Phi) \nonumber
\\
V(\Phi)&=&- \mu^2\Phi^\dagger \Phi + \lambda (\Phi^\dagger \Phi)^2
\;;\;\lambda>0\nonumber
\label{lsbs}
\end{eqnarray}
where,
\begin{eqnarray}
\Phi & = & \left(
\begin{array}{c}\phi^+ \\
 \phi_0 \end{array}\right)
\nonumber\\[2mm]
D_\mu \Phi &  = &  ( \partial_\mu - \frac{1}{2} i g
\vec{\sigma}\cdot\vec{W}_\mu
-\frac{1}{2} i g' B_\mu) \Phi \nonumber
\label{SMS}
\end{eqnarray}
Here $\Phi$ is a fundamental complex doublet with hypercharge
$Y(\Phi)=1$ and $V(\Phi)$ is the simplest renormalizable potential.
$\vec{W}_{\mu}$ and $B_{\mu}$ are the gauge fields of $SU(2)_{L}$
and
$U(1)_{Y}$ respectively and $g$ and $g'$ are the corresponding gauge
couplings.

It is interesting to notice the similarities with the
Ginzburg-Landau Theory. Depending on the sign of the mass parameter
($-\mu^2$), there are two possibilities for the v.e.v. $<0|\Phi|0>$ that
minimizes the potential $V(\Phi)$,
\begin{itemize}
\item[1)]
$(-\mu^2)>0$: The minimum is at:
\[ <0|\Phi|0>=0 \]
The vacuum is $\ws$ symmetric and therefore no symmetry breaking occurs.
\item[2)]
$(-\mu^2)<0$: The minimum is at:
\[|<0|\Phi|0>|=\left(
\begin{array}{c} 0 \\
 \frac{v}{\sqrt{2}} \end{array}\right)\;\;;\;\; {\rm arbitrary}\;\;
  arg\;\Phi \;\;;\;\;
v\equiv \sqrt{\frac{\mu^2}{\lambda}} \]

Therefore, there are infinite degenerate vacua corresponding to infinite
posssible values of $arg\;\Phi$. Either of these vacua is $\ws$
non-symmetric and $U(1)_{em}$ symmetric. The breaking $\ws
\rightarrow
U(1)_{em}$ occurs once a particular vacuum is chosen. As usual, the
simplest choice is taken:
\[ |<0|\Phi|0>| = \left(
\begin{array}{c} 0 \\
 \frac{v}{\sqrt{2}} \end{array}\right)\;;\; arg\; \Phi \equiv 0\;;\;
v \equiv \sqrt{\frac{\mu^2}{\lambda}}  \]
\end{itemize}
The two above symmetric and non-symmetric phases of
the Electroweak Theory are clearly similar to the two phases of the
ferromagnet that we have described within the Ginzburg Landau
Theory context. In the SM, the field $\Phi$ replaces the magnetization
$\vec{M}$ and the potential $V(\Phi)$ replaces $V(\vec{M})$. The SM
order papameter is, consequently, $<0|\Phi|0>$. In the symmetric phase,
$V(\Phi)$ is as in Fig.2a, whereas in the non-symmetric phase, it is as in
Fig.2b.

Another interesting aspect of the Higgs Mechanism, as we have already
mentioned, is that it preserves the total number of polarization
degrees. Let us make the counting in detail:
\begin{itemize}
\item[1)] {\bf Before SSB} \\
4 massless gauge bosons: $W^{\mu}_{1,2,3}, B^{\mu}$ \\
4 massless scalars: The 4 real components of $\Phi$: 
$\phi_1,\phi_2,\phi_3,\phi_4$ \\

Total number of polarization degrees $= 4\times 2 + 4 = 12$
\item[2)] {\bf After SSB} \\
3 massive gauge bosons: $W^{\pm},Z$ \\
1 massless gauge boson: $\gamma$ \\
1 massive scalar: $H$ \\

Total number of polarization degrees: $3\times 3+1\times 2+1=12$
\end{itemize}
Furthermore, it is important to realize that one more degree than needed
is introduced into the theory from the beginning. Three of the real
components of $\Phi$, or similarly
$\phi^{\pm} \equiv \frac{1}{\sqrt 2}(\phi_1 \mp i \phi_2)$ and
$\chi=\phi_3$,
are the needed would-be Goldstone bosons and the fourth one $\phi_4$ is
introduced just to complete the complex doublet. After the symmetry
breaking, this extra degree translates into the apparition in the
spectrum of an extra massive scalar particle ,
the Higgs boson particle $H$.
 

\section{Lagrangian of The Electroweak Theory II} 
 
 In order to get the particle spectra and the particle masses we first
rewrite  the full SM Lagrangian which is $\ws$ gauge
invariant:
\begin{eqnarray}
\cl_{SM}&=&
\cl_f+\cl_G+\cl_{SBS}+\cl_{YW}
\nonumber
\label{lsm}
\end{eqnarray}
where, $\cl_f$, and $\cl_G$ have been given previously and $\cl_{SBS}$ and 
$\cl_{YW}$ are the SBS and the Yukawa Lagrangians
respectively,  

\begin{center}
\framebox{
$ \cl_{SBS} = (D_\mu \Phi)^\dagger (D^\mu \Phi)
+ \mu^2\Phi^\dagger \Phi - \lambda (\Phi^\dagger \Phi)^2 $ }
\framebox{
$\cl_{YW}= \lambda_e\bar{l}_L\Phi e_R+
\lambda_u\bar{q}_L\widetilde{\Phi}u_R+\lambda_d\bar{q}_L\Phi d_R+h.c.+
2^{nd}\; {\rm and}\; 3^{rd}\;{\rm families} $ }
\end{center} 

Here,
 \begin{eqnarray}
l_L& = & \left ( \begin{array}{c}\nu_L \\ e_L \end{array}\right )\;;\;\;
q_L =  \left (\begin{array}{c}u_L \\ d_L \end{array}\right ) \nonumber
\\
\Phi & = & \left(
\begin{array}{c}\phi^+ \\
 \phi_0 \end{array}\right)\;;\;\;
\widetilde{\Phi}=i\sigma_2\Phi^*=\left(\begin{array}{c}
\phi^*_0\\-\phi^-\end{array}\right)\nonumber
\label{fields}
\end{eqnarray}
Notice that $\cl_{SBS}$ is needed to provide the $M_W$ and $M_Z$ masses and
$\cl_{YW}$ is needed to provide the $m_f$ masses.

The following steps summarize the procedure to get the spectrum from
$\cl_{SM}$:
\begin{itemize}
\item[1.-] A non-symmetric vacuum must be fixed. Let us choose, for
instance,
  \[<0|\Phi|0>=\left( \begin{array}{c} 0\\ \frac{v}{\sqrt{2}}
  \end{array}\right ) \]
\item[2.-] The physical spectrum is built by performing 'small
oscillations' around this vacuum. These are parametrized by,
\[\Phi(x)=\exp{\left ( i\frac{\vec{\xi}(x)\vec{\sigma}}{v}\right ) }
\left(\begin{array}{c}
0\\ \frac{v+H(x)}{\sqrt{2}}\end{array}\right ) \]
where $\vec{\xi}(x)$ and $H(x)$ are 'small' fields.
\item[3.-] In order to eliminate the unphysical fields $\vec{\xi}(x)$
we make the
following gauge transformations:
\begin{eqnarray}
\Phi'&=&U(\xi)\Phi=\left(\begin{array}{c}0\\ \frac{v+H}
{\sqrt{2}}\end{array}\right)\;;\;
U(\xi)=\exp\left(-i
\frac{\vec{\xi}\vec{\sigma}}{v}\right) \nonumber\\
l_L'&=&U(\xi)l_L\;;\;e_R'=e_R\;;\;q_L'=U(\xi)q_L
\;;\;u_R'=u_R\;;\;d_R'=d_R \nonumber\\
\left(\frac{\vec{\sigma}\cdot\vec{W}'_{\mu}}{2}\right)&=&
U(\xi)\left(\frac{\vec{\sigma}\cdot\vec{W}_{\mu}}{2}\right) U^{-1}(\xi)
-\frac{i}{g}(\partial_{\mu}U(\xi))U^{-1}(\xi)\;;\;B'_{\mu}=B_{\mu}\nonumber
\end{eqnarray}
\item[4.-] Finally, the weak eigenstates are rotated to the mass
eigenstates which define the physical gauge boson fields:
\begin{eqnarray}
W^\pm_\mu & = & \frac{W'^1_\mu \mp i W'^2_\mu}{\sqrt 2},
\nonumber\\[2mm] Z_\mu & = & c_w W'^3_\mu - s_w B'_\mu
\nonumber, \\[2mm] A_\mu & = & s_w W'^3_\mu + c_w B'_\mu,\nonumber
\end{eqnarray}
\end{itemize}
It is now straightforward to read the masses from the following
terms of $\cl_{SM}$:
\begin{eqnarray}
(D_{\mu}\Phi')^{\dagger}(D^{\mu}\Phi')&=&\left(\frac{g^2v^2}{4}\right)
W^+_{\mu}W^{\mu -}+\frac{1}{2}\left(\frac{(g^2+g'^2)v^2}{4}\right)
Z_{\mu}Z^{\mu}+...\nonumber \\
V(\Phi')&=&\frac{1}{2}(2\mu^2)H^2+...\nonumber \\
\cl_{YW}&=&
\left(\lambda_e \frac{v}{\sqrt{2}}\right)\bar{e}'_L e'_R+
\left(\lambda_u \frac{v}{\sqrt{2}}\right)\bar{u}'_L u'_R+
\left(\lambda_d \frac{v}{\sqrt{2}}\right)\bar{d}'_L d'_R+...\nonumber
\end{eqnarray}
and get finally the tree level predictions:
\begin{eqnarray}
M_W&=&\frac{gv}{2}\;;\;M_Z=\frac{\sqrt{g^2+g'^2}v}{2}\nonumber \\
M_H&=&\sqrt{2}\mu\nonumber\\
m_e&=&\lambda_e\frac{v}{\sqrt{2}}\;;\;
m_u=\lambda_u\frac{v}{\sqrt{2}}\;;\;
m_d=\lambda_d\frac{v}{\sqrt{2}}\;;...\nonumber
\label{tree}
\end{eqnarray}
where,
$$ v=\sqrt{\frac{\mu^2}{\lambda}} $$
Finally one can rewrite $\cl_{SBS}$ and $\cl_{YW}$, after the application of 
the Higgs Mechanism, in terms of the physical scalar fields, and get not
just the mass
terms but also the kinetic and interaction terms for 
 the Higgs sector,
 $$\cl_{SBS}+\cl_{YW} \rightarrow \cl_H^{\rm free}+\cl_H^{\rm int}+...$$
where,
\begin{center}
\framebox{
$\cl_H^{\rm free}=\frac{1}{2}\partial_{\mu}H \partial^{\mu}H-
\frac{1}{2}M_H^2H^2$  }
\end{center}
and,
\begin{center}
\framebox{
\shortstack{
$\cl_H^{\rm int}=
-\frac{M_H^2}{2v}H^3-\frac{M_H^2}{8v^2}H^4$ \\
$\;\;\;\;-\frac{m_f}{v}\overline{f}Hf $  \\
$\;\;\;\;+M_W^2W_{\mu}^+W^{\mu\;-}\left(1+\frac{2}{v}H+
\frac{1}{v^2}H^2\right)$\\
$\;\;\;\;+\frac{1}{2}M_Z^2Z_{\mu}Z^{\mu}\left(1+\frac{2}{v}H+
\frac{1}{v^2}H^2\right)$
   }
     }
\end{center}

Some comments are in order.
\begin{itemize}
\item[-] All masses are given in terms of a unique mass parameter $v$
and the couplings $g$, $g'$, $\lambda$, $\lambda_e$, etc..
\item[-] The interactions of $H$ with fermions and with gauge bosons are
proportional to the gauge couplings and to the corresponding particle
masses:
\[ f\bar{f}H\;:\;-i\frac{g}{2}\frac{m_f}{M_W}\;;\;\;\;
W^+_{\mu}W^-_{\nu}H\;:\;igM_Wg_{\mu\nu}\;;\;\;\;
Z_{\mu}Z_{\nu}H\;:\;\frac{ig}{c_w}M_Zg_{\mu\nu}\]
\item[-] The v.e.v. $v$ is determined experimentally form $\mu$-decay.
By identifying the predictions of the partial width $\Gamma(\mu
\rightarrow
\nu_{\mu}
\bar{\nu}_e e)$ in the SM to low energies ($q^2<<M_W^2$) and in the V-A
Theory one gets,
\[\frac{G_F}{\sqrt{2}}=\frac{g^2}{8 M_W^2}=\frac{1}{2 v^2} \]
And from here,
\[v=(\sqrt{2} G_F)^{-\frac{1}{2}}= 246\;GeV\]
\item[-] The values of $M_W$ and $M_Z$ were anticipated successfully
quite before they were measured in experiment. The input
parameters were $\theta_w$, the fine structure constant $\alpha$ and
$G_F$. Before LEP these were the best measured electroweak parameters.
\item[-] In contrast to the gauge boson sector, the Higgs boson mass
$M_H$ and the Higgs self-coupling $\lambda$ are completely undetermined
in the SM. They are related at tree level by, $\lambda=\frac{M_H^2}{2v^2}$.
\item[-] The hierarchy in the fermion masses is also
completely undetermined in the SM.
\end{itemize}

\section{Theoretical Bounds on $M_H$}

In this section we summarize the present bounds on $M_H$ from the
requirement of consistency of the theory.

\subsection{ Upper bound on $M_H$ from Unitarity}

Unitarity of the scattering matrix together with the elastic
approximation for the total cross-section and the Optical Theorem
imply certain elastic unitarity conditions for the partial wave
amplitudes. These, in turn, when applied in the SM to scattering
processes involving the Higgs particle, imply an upper limit on the
Higgs mass. Let us see this in more detail for the simplest case of
scattering of massless scalar particles: $1+2\rightarrow 1+2$.

The decomposition  of the amplitude in terms of partial waves is given
by:
\begin{eqnarray}
T(s,\cos\theta)&=&16\pi\sum_{J=0}^{\infty}(2J+1)a_J(s)P_J(\cos\theta)
\nonumber
\end{eqnarray}
where $P_J$ are the Legendre polynomials.

The corresponding differential cross-section is given by:
\begin{eqnarray}
\frac{d\sigma}{d\Omega}&=&\frac{1}{64\pi^2s}|T|^2
\nonumber
\end{eqnarray}

Thus, the elastic cross-section is written in terms of partial
waves as:
\begin{eqnarray}
\sigma_{\rm el}&=&\frac{16\pi}{s}\sum_{J=0}^{\infty}(2J+1)|a_J(s)|^2
\nonumber
\label{elast}
\end{eqnarray}
On the other hand, the Optical Theorem relates the total cross-section
with the forward elastic scattering amplitude:
\begin{eqnarray}
\sigma_{\rm tot}(1+2\rightarrow {\rm anything})&=&\frac{1}{s}
 Im\;T(s,\cos\theta=1) \nonumber
\label{tot}
\end{eqnarray}
In the elastic approximation for $\sigma_{\rm tot}$ one gets
$\sigma_{\rm tot}\approx \sigma_{\rm el}$.
From this one finally finds,
\begin{eqnarray}
Im\;a_J(s)&=&|a_J(s)|^2\;;\;\forall J \nonumber
\label{part}
\end{eqnarray}
This is called the elastic unitariry condition for partial wave
amplitudes. It is easy to get from this the following
inequalities:
\begin{eqnarray}
& &|a_J|^2\leq 1\;;\;0\leq Im\;a_J\leq 1\;;\;|Re\; a_J|\leq \frac{1}{2}
\;;\; \forall J \nonumber
\end{eqnarray}
These are necessary but not sufficient conditions for elastic unitarity.
It implies that if any of them are not fulfiled then the elastic
unitarity
condition  also fails, in which case the unitarity of
the theory is said to be violated.

Let us now study the particular case of $W_L^+W_L^-$ scattering in the
SM and find its unitarity conditions. The $J=0$ partial wave can be
computed from:
\begin{eqnarray}
a_0(W_L^+W_L^-\rightarrow W_L^+W_L^-)&=&\frac{1}{32\pi}\int_{-1}^1
T(s,\cos\theta)d(\cos\theta)\nonumber
\end{eqnarray}
where the amplitude $T(s,\cos \theta)$ is given  by,
\begin{eqnarray}
T(W_L^+W_L^-\rightarrow W_L^+W_L^-)&=&
-\frac{1}{v^2}\{-s-t+\frac{s^2}{s-M_H^2}+\frac{t^2}{t-M_H^2}
      +2M_Z^2+
\nonumber \\ & &
      \frac{2M_Z^2s}{t-M_Z^2}+
\frac{2t}{s}(M_Z^2-4M_W^2)-\frac{8s_W^2M_W^2M_Z^2s}{t(t-M_Z^2)}\}
\nonumber 
\end{eqnarray}
 By
studying the large energy limit  of
$a_0$ one finds,
\begin{eqnarray}
|a_0|& \stackrel{s>>M_H^2,M_V^2}{\longrightarrow} &\frac{M_H^2}{8\pi v^2}
\nonumber
\end{eqnarray}
Finally, by requiring the unitarity condition $|Re\;a_0|\leq \frac{1}{2}$
one gets the following upper bound on the Higgs
mass:
\begin{eqnarray}
M_H<860\; GeV
\nonumber
\end{eqnarray}
One can repeat the same reasoning for different channels and find
similar or even tighter bounds than this one \cite{unitarity,screen}.

At this point, it should be mentioned that these upper bounds based on
perturbative unitarity do not mean that the Higgs particle cannot be
heavier than these values. The conclusion should be, instead, that for
those large $M_H$ values a perturbative approach is not valid and
non-perturbative techniques are required. In that case, the Higgs
self-interactions governed by the coupling $\lambda$ become strong and new
physics phenomena may appear in the $O(1\;TeV)$ range. In particular, 
the scattering of longitudinal gauge bosons may also become strong in that
range \cite{unitarity} and behave similarly to what happens in $\pi\pi$
scattering in the $O(1\;GeV)$ range. Namely, there could appear new
resonances, as it occurs typically in a theory with strong interactions. 
This new interesting phenomena could be studied at the next hadron 
collider, LHC \cite{unitarity,terron}.    

\subsection { Upper bound on $M_H$ from Triviality}

Triviality in $\lambda \Phi^4$ theories  \cite{triv} (as, for instance,
the scalar
sector of the SM) means that the particular value of the renormalized
coupling of $\lambda_R=0$ is the unique fixed point of the theory. A
theory with
$\lambda_R=0$
contains non-interacting particles and therefore it is trivial. This
behaviour can already be seen in the renormalized coupling at one-loop
level:
\begin{eqnarray}
\lambda_R(Q)&=&\frac{\lambda_0}{1-\frac{3}{2\pi^2}\lambda_0
\log(\frac{Q}{\Lambda})}\;;\;\lambda_0\equiv \lambda_R(Q=\Lambda) \nonumber
\end{eqnarray}
As we attempt to remove the cut-off $\Lambda$ by taking the limit
$\Lambda \rightarrow \infty$ while $\lambda_0$ is kept fixed to an
arbitrary but finite value, we find out that
$\lambda_R(Q)\rightarrow 0$ at any finite energy value $Q$. This, on the
other hand, can be seen as a consequence of the existence of the well
known Landau pole of $\lambda \Phi^4$ theories.

The trivilaty of the SBS of the SM is cumbersome since we need a
self-interacting scalar system to generate $M_W$ and $M_Z$ by the Higgs
Mechanism. The way out from this apparent problem is to assume that the
Higgs potential $V(\Phi)$ is valid just below certain 'physical' cut-off
$\Lambda_{\rm phys}$. Then, $V(\Phi)$ describes an effective low energy
theory which emerges from some (so far unknown) fundamental physics with
$\Lambda_{\rm phys}$ being its characteristics energy scale. We are
going to see next that this assumption implies an upper bound on 
$M_H$ \cite{1loop}.

Let us assume some concrete renormalization of the SM parameters. The
conclusion does not depend on this particular choice. Let us define, for
instance, the renormalized Higgs mass parameter as:
\begin{eqnarray}
M_H^2 & = &2 \lambda_R(v)v^2 \nonumber
\label{mh}
\end{eqnarray}
where,
\begin{eqnarray}
\lambda_R(v)&=&\frac{\lambda_0}{1-\frac{3}{2\pi^2}\lambda_0
\log(\frac{v}{\Lambda_{\rm phys}})}\nonumber
\label{lr}
\end{eqnarray}
Now, if we want $V(\Phi)$ to be a sensible effective theory, we must
keep all the renormalized masses below the cut-off and, in particular,
$M_H<\Lambda_{\rm phys}$. However,  one
 can see that for arbitrary values of $\Lambda_{\rm phys}$ it is not
always possible. By increasing the value of $\Lambda_{\rm phys}$, $M_H$
decreases and the other way around, by lowering $\Lambda_{\rm phys}$,
$M_H$ grows. There is a crossing point where $M_H\approx \Lambda_{\rm
phys}$ which happens to be around an energy scale of approximately
$1\;TeV$. Since we want to keep the Higgs mass below the physical
cut-off, it implies finally the announced upper bound,
\[M_H^{\rm 1-loop}< 1\;TeV\]
Of course, this should be taken just as a perturbative estimate of the
true triviality bound. A more realistic limit must come from a
non-perturbative treatment. In particular, the analyses performed on
the lattice  \cite{lattice} confirm this behaviour and place even tighter
limits. The following bound is found,
\[ M_H^{\rm Lattice} < 640 \;GeV \]
Finally, a different but related perturbative upper limit on $M_H$ can
be found by analysing the renormalization group equations in the SM to
one-loop. Here one includes, the scalar sector, the gauge boson
sector and restricts the fermionic sector to the third generation.  By
requiring the theory to be perturbative (i.e. all the couplings be
sufficiently small) at all the energy scales below some fixed high
energy, one finds a maximum allowed $M_H$ value \cite{Cabetal}.
For instance, by fixing
this energy scale to $10^{16}\;GeV$ and for $m_t=170\;GeV$ one gets:
\[ M_H^{\rm RGE} < 170\;GeV \]
Of course to believe in perturbativity up to very high energies
could be just a theoretical prejudice. The existence
of a non-perturbative regime for the scalar sector of the SM is still a
possibility and one should be open to new proposals in this concern.

\subsection { Lower bound on $M_H$ from Vacuum Stability}

 Once the asymmetric vacuum of the $\ws$ theory has been fixed, one must
require
this vacuum to be stable under quantum corrections. In principle,
quantum corrections could destabilize the asymmetric vacuum and change
it to the symmetric one where the SSB does
not take place. This phenomenom can be better explained in terms of the
effective potential with quantum corrections included. Let us
take, for instance, the effective potential of the Electroweak Theory
to one loop in the small $\lambda$ limit:
\begin{eqnarray}
V_{\rm eff}^{\rm 1-loop}(\Phi) &\simeq & -\mu^2\Phi^{\dagger}\Phi+
\lambda_R(Q_0) (\Phi^{\dagger}\Phi)^2+\beta_{\lambda}
(\Phi^+\Phi)^2\log\left ( \frac{\Phi^{\dagger}\Phi}{Q_0^2} \right )
\nonumber
\end{eqnarray}
where, $\beta_{\lambda} \equiv \frac{d\lambda}{dt} \simeq 
 \frac{1}{16\pi^2}\left [
-3\lambda_t^4+\frac{3}{16}(2g^4+(g^2+g'^2)^2)\right ]$.

The condition of extremum is:
\begin{eqnarray}
\frac{\delta V_{\rm eff}^{\rm 1-loop}}{\delta\Phi}&=&0
\nonumber
\end{eqnarray}
which leads to two possible solutions: a) The trivial vacuum with
$\Phi=0$; and b) The non-trivial vacuum with $\Phi=\Phi_{\rm vac}\neq
0$. If we want the true vacuum to be the non-trivial one we
must have:
\begin{eqnarray}
V_{\rm eff}^{\rm 1-loop}(\Phi_{\rm vac})& < &
V_{\rm eff}^{\rm 1-loop}(0)
\nonumber
\label{stab}
\end{eqnarray}
However, the value of the potential at the minimum depends on the size
of its second derivative:
\begin{eqnarray}
& & M_H^2\equiv \frac{1}{2}\{\frac{\delta^2V}
{\delta\Phi^2} \}_{\Phi=\Phi_{\rm vac}}
\nonumber
\end{eqnarray}
and, it turns out that for too low values of $M_H^2$ the condition
above  turns over. That is, $V(0)<V(\Phi_{\rm vac})$
and the true vacuum changes to the trivial one. The condition for vacuum
stability then implies a lower bound on $M_H$ \cite{vstab1}. More
precisely,
\begin{eqnarray}
M_H^2 &> &\frac{3}{16\pi^2v^2}(2M_W^4+M_Z^4-4m_t^4)
\nonumber
\end{eqnarray}
Surprisingly, for $m_t>78\;GeV$ this bound dissapears and, moreover,
$V_{\rm eff}^{\rm 1-loop}$
becomes unbounded from below!. Apparently it seems a
disaster since the top mass is known at present and is certainly larger
than this value. The solution to this problem relies in the fact that
for such input values, the 1-loop approach becomes unrealistic and a
2-loop analysis of the effective potential is needed. Recent studies
indicate that by requiring vacuum stability at 2-loop level and up to
very large energies of the order of $10^{16}\;GeV$, the following lower
bound is found \cite{vstab2}:
\begin{eqnarray}
M_H^{\rm v.stab.} &> &132\;GeV \nonumber
\end{eqnarray}
This is for $m_t=170\; GeV$ and $\alpha_s=0.117$ and there is an
uncertainty in this bound of $5$ to $10$ $GeV$ from the uncertainty in
the $m_t$ and $\alpha_s$ values.

\section{SM predictions}

In the following we present the tree level predictions from the 
SM and compare them with the present experimental values. 
The experimental values presented here 
 (unless explicitely stated otherwise)  have been borrowed from the talk 
by D.Karlen given at the ICHEP'98  Vancouver Conference \cite{status}. 
For a more 
detailed discussion on the experimental tests of the SM see the
lectures of L. Nodulman.

\subsection{Gauge Boson Masses}
Before the discovery of the $W^{\pm}$, $Z$ gauge bosons, the best known SM
parameters were $\alpha$, $G_F$ and $sin^2\theta_w$. The present values are
highly precise:
$$\alpha^{-1}_{exp}=137.0359895\pm 0.0000061$$
from atomic, molecular and nuclear data, and
$$G_F^{exp}=(1.16639\pm 0.000022)\times 10^{-5}\;GeV^{-2}$$
from $\mu-$decay.\\
$sin^2\theta_w$ was measured firstly in the seventies in $\nu N$ scattering
experiments. The ratio of the cross-sections
for neutral currents and charged currents as predicted in the SM is a
known function of $sin\theta_w$,
$$\frac{\sigma_{NC}(\nu q\rightarrow \nu q)}  
 {\sigma_{CC}(\nu q\rightarrow l q')}\sim f(sin^2\theta_w) $$
It is the measurement of this ratio what provides a measurement of
 $sin^2\theta_w$. The present experimental value is,  
 $$sin^2\theta_w|_{\rm exp}=0.2255\pm 0.0021$$
The SM does not predict a numerical value for $M_W$ and $M_Z$ but provides some
relations among the relevant parameters. These relations are different to tree
level than to, for instance, one-loop level. In particular, the following
relations hold to tree level in the SM,
$$M_W=\frac{gv}{2}\;;\frac{G_F}{\sqrt{2}}=\frac{g^2}{8M_W^2}=\frac{1}{2v^2}
\;;\;g=\frac{e}{s_w}\;;\;\rho_{SM}^{\rm tree}\equiv\frac{M_W^2}{M_Z^2c_w^2}=1$$
From these expressions it is inmediate  to derive the two following
tree level relations,
$$M_W=\left(\frac{\pi\alpha}{G_F\sqrt{2}}\right)^{\frac{1}{2}}
\frac{1}{sin\theta_w}$$
$$M_Z=\left(\frac{\pi\alpha}{G_F\sqrt{2}}\right)^{\frac{1}{2}}
\frac{1}{sin\theta_w cos\theta_w}$$
Finally, by  inserting the experimental values of $\alpha$, $G_F$ and 
$\theta_w$ into these expressions one gets the
tree level values for the gauge boson masses,
$$M_W^{\rm tree}=78\; GeV\;\;;\;\;M_Z^{\rm tree}=89\;GeV$$
The discovery of the $W^\pm$ and $Z$ gauge bosons in 1983 at the CERN 
$SpS$ collider \cite{SpS} lead to the definitive confirmation of the 
validity of the SM. Notice that the measured masses were surprisingly 
close to the
SM tree level predictions,
$$M_W^{\rm SpS}=(81\pm 2)\;GeV\;\;;\;\; M_Z^{\rm SpS}=(93\pm 3)\;GeV$$  
The present experimental values are  very precise,
\begin{eqnarray}
M_W^{\rm exp}&=&(80.41\pm 0.09)\;GeV\;(p\overline{p}) \nonumber \\
 & &(80.37\pm 0.09)\;GeV\;({\rm LEP})\nonumber
 \end{eqnarray} 
\begin{eqnarray}
M_Z^{\rm exp}&=&(91.1867\pm0.0021)\;GeV\;({\rm LEP}) \nonumber
\end{eqnarray} 

\subsection{Gauge Boson Decays}
The $W^\pm$ and $Z$ gauge bosons can decay either in quarks or in leptons 
within the SM. The dominant decays are clearly into quarks due to the  
extra color factor, $N_C$, which is not present in the leptonic decays.
The tree level predictions for the partial widths in the approximation of 
neglecting the fermion masses are the following,
$$\Gamma(W^+\rightarrow e^+\nu_e)=\frac{g^2}{48\pi}M_W=
\frac{G_FM_W^3}{6\sqrt{2}\pi}=0.232\;GeV$$
$$\Gamma(W^+\rightarrow \mu^+\nu_{\mu})=\Gamma(W^+\rightarrow\tau^+\nu_{\tau}) 
=\Gamma(W^+\rightarrow e^+\nu_e)$$
$$\Gamma(W^+\rightarrow u_i\overline{d}_j)=
N_C |U_{ij}|^2\frac{G_FM_W^3}{6\sqrt{2}\pi}=0.232\;N_C|U_{ij}|^2\;GeV$$
$$\Gamma(Z\rightarrow f\overline{f})=
\kappa_f\frac{G_FM_Z^3}{6\sqrt{2}\pi}(g_{Vf}^2+g_{Af}^2)=
0.3318 \kappa_f(g_{Vf}^2+g_{Af}^2)\;GeV$$
where, $U_{ij}$ are the CKM matrix elements and,
$$\kappa_f=1\;,\; f=l,\nu \;;\;\kappa_f=N_C\;,\;
f=q\;;\;g_{Vf}=T^f_3-2Q^fs_w^2\;;\;g_{Af}=T^f_3$$
 
In Table 3 it is shown the tree level predictions in the SM, for the total 
$Z$ and $W^{\pm}$ widths, $\Gamma_Z$ and $\Gamma_W$,  and the ratios:
$$R_e=\frac{\Gamma(Z\rightarrow {\rm hadrons})}{\Gamma_e}\;;\;
  R_b=\frac{\Gamma(Z\rightarrow b\overline{b})}
  {\Gamma(Z\rightarrow {\rm hadrons}) }\;;\;
  R_c=\frac{\Gamma(Z\rightarrow c\overline{c})}
  {\Gamma(Z\rightarrow {\rm hadrons}) } $$
The numerical predictions shown here use as input  the present
experimental values for $M_Z$, $G_F$, $\alpha$ and $sin^2\theta_w$,
and the $M_W$ value that one gets with those experimental values put into 
the previous SM 
tree level relation. That is, $M_W=80.94\;GeV$.   
\begin{table}[htb]
\begin{center}
\caption{Confronting tree level SM predictions with data}
\begin{tabular}{lrr}
\hline
Parameter & Tree Level SM & Exp. value\\
\hline
$\Gamma_Z(GeV)$ & $2.474$ & $2.4948\pm 0.0025$   \\
$\Gamma_W(GeV)$ & $2.09$  & $2.06\pm 0.06$       \\
$R_e$           & $20.29$ & $20.765\pm 0.026$    \\
$R_b$           & $0.219$ & $0.21656\pm 0.00074$ \\
$R_c$           & $0.172$ & $0.1733\pm 0.0044$\\
\hline
\end{tabular}
\end{center}
\end{table} 

By comparing the tree level SM results with the present experimental values 
it is clear that they provide reasonable good predictions. However, due to the
high level of precision of the present measurements one can also conclude from
this table that some of the tree level predictions are  already not compatible
with data. In fact in some observables they are out by 
several standard deviations. This is a clear indication that the SM radiative
corrections must be included in the theoretical predictions \cite{Hollik}. 
The present
experimental analysis of the SM parameters, in fact, do include
these radiative corrections.  The summary of measurements included in the
combined analysis of SM parameters from LEP and SLC can be found in Nodulman
lectures.
   
\subsection{Top Quark Physics}

As has been shown before, the top mass $m_t$ is not predicted in the SM.
Instead, the
SM provides, via The Higgs Mechanism,  the tree level relation,
$$m_t=\lambda_t \frac{v}{\sqrt{2}}=\lambda_t \left(\frac{1}{2\sqrt{2}G_F}
\right)^{\frac{1}{2}}$$ 
which gives $m_t$ in terms of the top Yukawa coupling $\lambda_t$. But, 
$\lambda_t$ as the other fermion Yukawa couplings are unknown parameters in
the SM. 

\begin{figure}
\hspace{1.5cm}
\epsfysize8cm
\centerline{\epsffile{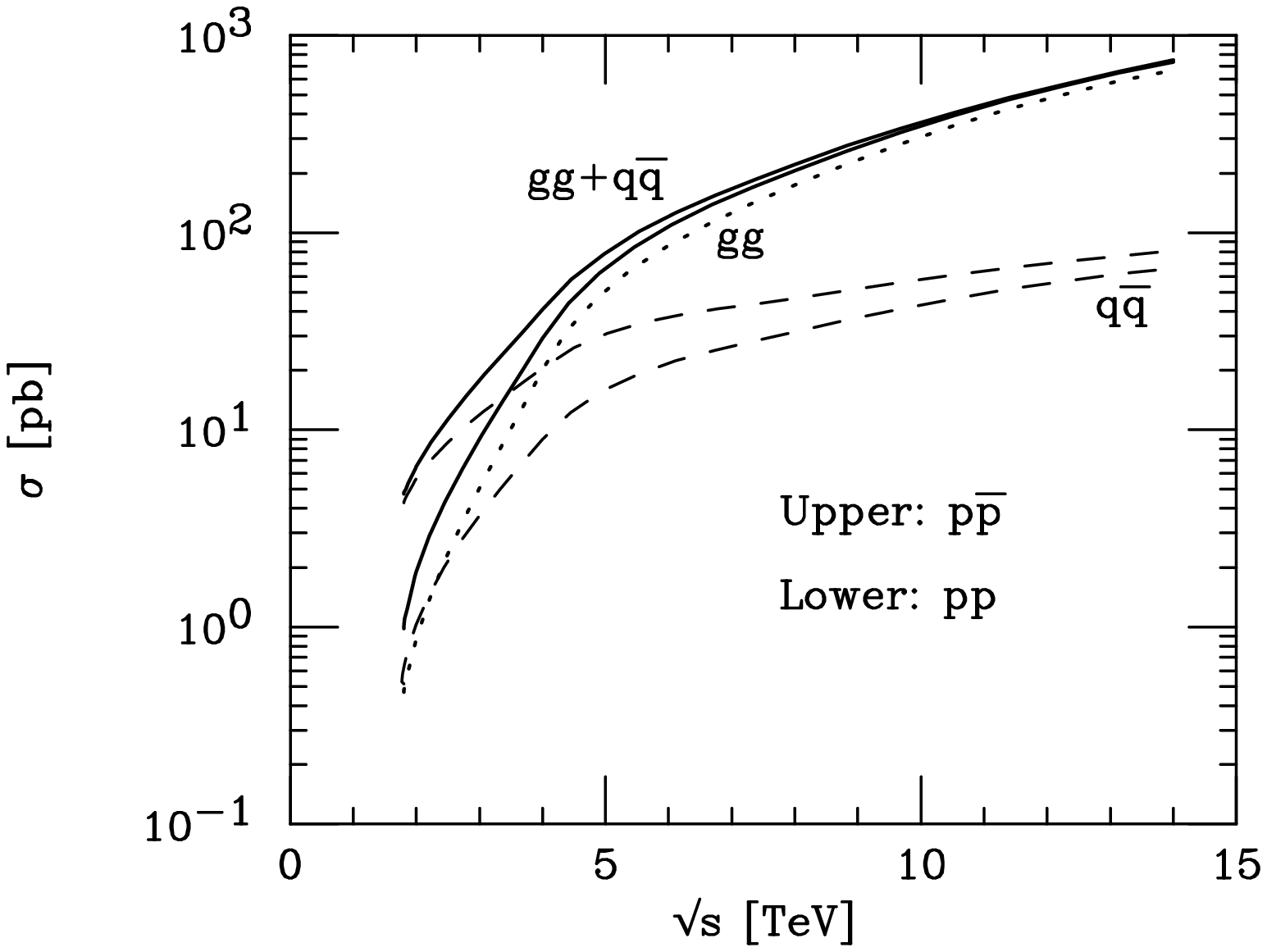}}
\caption{$t\overline{t}$ production at hadron colliders}
\end{figure}

The existence of the top quark, however, was never questioned seriously, since 
there were several strong arguments supporting the need of this third generation
fermion. On one hand, the top quark was needed to avoid unwanted flavour
changing neutral currents (FCNC). On the other hand the top quark was needed to
avoid unwanted $\ws$ anomalies. Thus,      
The top quark was expected for a long time. Its discovery finally occured in
1994
at the 
$p\overline{p}$ TeVatron collider at Fermilab \cite{top}. 
 The present experimental value 
of the top mass as provided by the two Tevatron experiments is,
$$m_t^{\rm exp}= (173.8 \pm 5.0)\;GeV\;({\rm CDF}+{\rm D}0) $$ 
It is remarkable this much larger mass value than the rest of the fermion
masses. In fact, for this top mass value one can extract the corresponding 
Yukawa coupling and get $\lambda_t \sim 1$ which is a rather large value,
although it can still be considered as a perturbative coupling. To the
question, why the top quark is so heavy?, there is no answer within the SM.

Concerning the top quark decays, the dominant one is by far the decay into a
$W^+$ gauge boson and a bottom quark. The SM tree level prediction for the
partial width, in the
approximation of neglecting $m_b$, is  
$$\Gamma(t\rightarrow W^+b)=\frac{G_Fm_t^3}{8\pi\sqrt{2}}
|U_{tb}|^2\left(1-\frac{M_W^2}{m_t^2}\right) 
\left(1+2\frac{M_W^2}{m_t^2}\right) \sim 2\;GeV$$
There is not experimental meassurement of the total  or partial width yet.
 
Regarding the top production, it is obvious that, given the large
mass value, it  can only be produced at present in the TeVatron collider. The
future hadron collider LHC at CERN will provide additional interesting
information on top quark physics. In Figure 3 it is shown the cross-section for 
$t\overline{t}$ production at TeVatron and LHC from the various possible
channels. It is clear from the figure that
 $q\overline{q}\rightarrow t\overline{t}$
 is the dominant process at TeVatron, whereas 
$gg\rightarrow t\overline{t}$ will dominate at LHC \cite{topreview}. 

\subsection{Higgs Physics}

The Higgs mass $M_H$ is not predicted in the SM either. The Higgs Mechanism 
provides $M_H$ as a function of the Higgs self-coupling $\lambda$ and
$v=246\;GeV$,
$$M_H=\sqrt{2}\mu=\sqrt{2v^2\lambda}$$  
but, $\lambda$ is also unknown. Therefore, $M_H$ can take any
value in the SM. 
As we have seen, the unique restrictions on $M_H$ come from the consistency
of the theory, that is
from unitarity, triviality, and vacuum stability arguments. 
 
\begin{figure}
\epsfysize=20cm
\vspace{-8cm}
\centerline{\epsffile{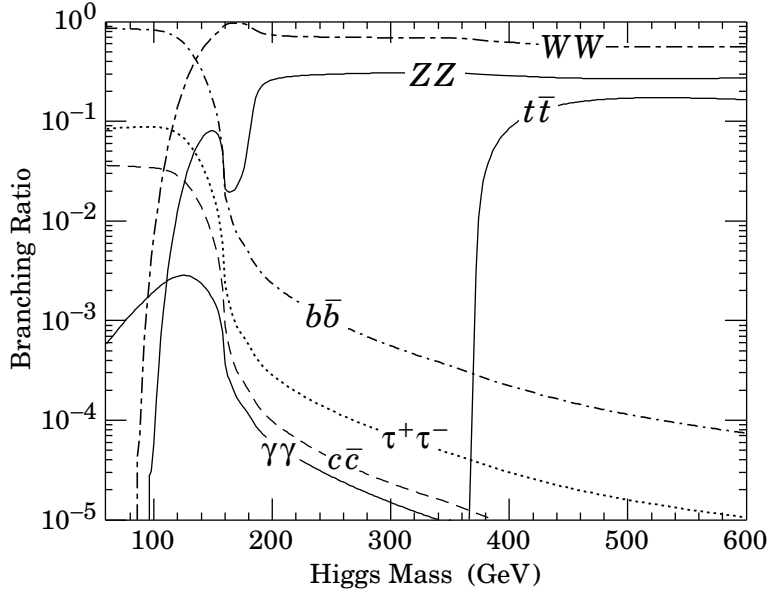}}
\vspace{-3cm}
\caption{ Branching ratios for Higgs decays as a function of $M_H$}
\end{figure}

Contrary to the top quark case, there are no strong theoretical arguments
(anomalies, etc)
supporting the need of this scalar elementary particle, $H$. 

Regarding the Higgs decays, the tree level SM predictions for the partial 
widths are the following,
$$\Gamma (H\rightarrow W^+W^-)=\frac{G_FM_H^3}{8\sqrt{2}\pi}
\beta_W\left(\beta_W^2+\frac{12M_W^4}{M_H^4}\right)$$
$$\Gamma (H\rightarrow ZZ )=\frac{G_FM_H^3}{16\sqrt{2}\pi}
\beta_Z\left(\beta_Z^2+\frac{12M_Z^4}{M_H^4}\right)$$ 
$$\Gamma(H\rightarrow f\overline{f})=\frac{G_Fm_f^2M_H}{4\sqrt{2}\pi}
\beta_f^3\xi$$
where,
$$\beta_P=\sqrt{1-4\frac{m_P^2}{M_H^2}}\;,\;P=W^{\pm},Z,f\;\;;\;\;
\xi=3\;{\rm if}\;f=q\;,\;\xi=1\;{\rm if}\;f=l$$
 
In Figure 4 there are shown the Higgs branching ratios as a function of the Higgs
mass. For low Higgs mass the dominant decay is to $b\overline{b}$. Above the
weak gauge bosons threshold, the dominant decay is to $W^+W^-$.

The total Higgs width ranges from very small values for the low $M_H$ region
to very large values in the high $M_H$ region. Some examples are,
$$\Gamma_H^{\rm tot}(M_H=100\;GeV)\sim 5\times10^{-3}\;GeV\;;\;
\Gamma_H^{\rm tot}(M_H=1000\;GeV)\sim 570\;GeV$$

\section{Experimental bounds on $M_H$}

The search of the Higgs particle at present $e^+e^-$ and
$\bar{p}p$ colliders is a rather difficult task due the smallness of the
cross-sections for Higgs production  which, in turn, is
explained
in terms of the small couplings of the Higgs particle to light fermions:
$H\bar{f}f\leftrightarrow -i\frac{g}{2}\frac{m_f}{M_W}$. On the other
hand, at present available energies, the dominant decay channel is
$H\rightarrow b\bar{b}$  which is not easy to study due to
the complexity of the final state and the presence of large 
backgrounds \cite{hhg}.
 
\subsection{ Higgs search at $e^+e^-$ colliders (LEP, SLC)}

The Higgs search  during the first period of LEP (LEPI) and SLC
 was done mainly by
analysing the process:
\[ e^+e^- \rightarrow Z \rightarrow Z^*H\]
with the virtual $Z^*$ decaying as $Z^*\rightarrow 
l^+l^-,\;\nu \overline{\nu},\;q\overline{q}$ 
 and the Higgs particle decaying as
$H\rightarrow b\bar{b}$.

At LEPI with a center-of-mass-energy adjusted to the $Z$ mass,
$\sqrt{s}\sim
M_Z$, a very high statistic was reached and a systematic
search of the Higgs particle for all kinematically allowed $M_H$ values
was possible. The absence of any experimental signal from the Higgs
particle implied a lower bound on $M_H$. The last reported bound
from LEPI was:
\[ M_H > 66\; GeV \; (95\% C.L.)\;\;(LEPI)  \]
In the second phase of LEP, LEPII, a center-of-mass-energy of up to
$\sqrt{s}\sim 189 GeV$ is at present reached. The relevant
process for Higgs searches is:
\[ e^+e^-\rightarrow Z^* \rightarrow ZH\]
where now the intermediate $Z$ boson is virtual and the final $Z$ is on
its mass shell. The analyses of the various relevant $Z$ and $H$ decays
at LEPII 
give at present the following averaged lower Higgs mass bound:
\[ M_H>89.8\;GeV \;(95\%C.L)\;\;(LEPII)  \]
In addition to the direct bounds on $M_H$ from LEP and SLC data, a great effort
is being done also in the search of indirect Higgs signals from its
contribution to electroweak quantum corrections. In fact, there are
 already  interesting upper experimental bounds on $M_H$
from the
meassurement of observables as $\Delta \rho$, $\Delta r$ and other related ones
 whose prediction in the SM
is well known. It is interesting to mention that neither the Higgs
particle nor the top quark decouple from these low energy observables.
It means that the quantum effects of a virtual $H$ or $t$ do not vanish
in the asymptotic limit of infinitely large $M_H$ or $m_t$ respectively. For
instance, the leading corrections to $\Delta \rho$, and $\Delta r$ in
the large $m_t$ and large $M_H$ limits are respectively:
\begin{eqnarray}
(\Delta\rho)_t &=& \frac{\sqrt{2}G_F3}{16\pi^2}m_t^2+...
\nonumber \\
(\Delta\rho)_H&=&- \frac{\sqrt{2}G_FM_W^2}{16\pi^2}3\frac{s_w^2}{c_w^2}
\left(\log\frac{M_H^2}{M_W^2}-\frac{5}{6}\right)+...
\nonumber \\
(\Delta r)_t &=&-\frac{c_w^2}{s_w^2}\frac{\sqrt{2}G_F3}{16\pi^2}m_t^2+...
\nonumber \\ 
(\Delta r)_H &=&\frac{\sqrt{2}G_FM_W^2}{16\pi^2}
\frac{11}{3}\left(\log\frac{M_H^2}{M_W^2}-\frac{5}{6}\right)+...
\nonumber
\end{eqnarray}

Whereas the top corrections grow with the mass as $m_t^2$, the Higgs
corrections are milder growing as $\log M_H^2$. It means that the top
non-decoupling effects at LEP are important. In fact they have been
crucial in the search of the top quark and have provided one of the
first indirect indications of the 'preference of data' for large $m_t$
values. This helped in the search and final  discovery of the
top quark at TeVatron.

The fact that the Higgs non-decoupling effects are mild  was announced a
long time ago by T.Veltman in the so-called 
{\it Screening Theorem} \cite{screen}.
  This theorem states that, {\it at one-loop, the dominant
quantum
corrections from a heavy Higgs particle to electroweak observables grow,
at most, as} $\log M_H$. {\it The Higgs corrections are of the generic
form}:
\[g^2(\log\frac{M_H^2}{M_W^2}+g^2\frac{M_H^2}{M_W^2}+...)\]
{\it and the potentially large effects proportional to} $M_H^2$ {\it are
'screened' by additional small} $g^2$ {\it factors}.
 
The present global analysis of SM parameters at LEP and SLC with the radiative
 corrections
included, gives the following  upper Higgs mass bound:
$$M_H<280\;GeV\;(95\%\;C.L.)\;\;(LEP+SLC)$$ 

\subsection{ Higgs search at hadronic colliders }

The relevant subprocesses for Higgs production at hadronic $pp$ and
$p\bar{p}$ colliders are: gluon-gluon fusion ($gg\rightarrow H$), $WW$ and
$ZZ$ fusion ($qq\rightarrow qqH$), $t\overline{t}$ fusion 
($gg\rightarrow t\overline{t}H$) and $Z$ ($W$) bremsstrahlung 
($q\overline{q}\rightarrow Z(W)H$).
 
 At present available energies the dominant subprocess is $gg$-fusion. This is 
 the case of 
TeVatron with a center-of-mass-energy of $\sqrt{s}=1.8\;TeV$ and an 
integrated luminosity
of $L= 100\;pb^{-1}$ per experiment.
 However, due to the
cleanest signature of the $Z$ and $W$ bremsstrahlung subprocesses and the fact
that this
has less background, this channel is the most studied one at TeVatron.
The Higgs searches at TeVatron have not provided yet any lower Higgs mass
bound. The sensitivity of the present search \cite{tevatron}
 is limited by statistics to
a cross section approximately two orders of magnitude larger than the predicted
cross section for SM Higgs production. For the next TeVatron run there will be 
a twenty-fold increase in the total integrated luminosity per experiment.
However, it   
is still insufficient to reach say a $120\;GeV$ Higgs mass, unless the total
detection efficiency be improved by one order of magnitude. The viability of
this improvement is at present under study.

The Higgs search at the LHC collider being built at
CERN, is very promising. In particular, it will cover the whole Higgs mass
range and hopefully will be able to distinguish between the various
possibilities for the SBS of the Electroweak Theory. 
For a review on Higgs searches at LHC
see F. Pauss lectures.

\section{Acknowledgements}

I thank Tom Ferbel for the invitation to this interesting and enjoyable school
and for the perfect organization. I am grateful to all the students and
lecturers for making many interesting questions and comments 
during my lectures and for keeping a very pleasant atmosphere for fruitful 
discussions. I have also profited from discussions with J.Troc\'oniz. I wish
to thank C.Glasman, P.Seoane and G.Yepes for their help with the figures. 
This work was supported in part by the Spanish CICYT 
  under project AEN97-1678.   


\end{document}